\documentclass[11pt,a4paper]{article}

\usepackage{jcappub}

\usepackage[english]{babel}
\usepackage{amsmath,amsfonts,amssymb}
\usepackage{comment}
\usepackage{bbold}
\usepackage{color}
\usepackage{slashed}

\usepackage[top=30mm,bottom=30mm,left=20mm,right=20mm]{geometry}

\usepackage{hyperref}
\usepackage{appendix}

\usepackage{booktabs}
\usepackage{graphicx}
\usepackage[font=normalsize]{caption}
\usepackage{ulem}
\usepackage{enumitem}

\usepackage{multirow}
\usepackage{arydshln}

\newcommand{\SU}[0]{\mathrm{SU}}
\newcommand{\U}[0]{\mathrm{U}}
\newcommand{\YU}[0]{\mathbf{Y}_u}
\newcommand{\YD}[0]{\mathbf{Y}_d}
\newcommand{\YE}[0]{\mathbf{Y}_e}
\newcommand{\YNU}[0]{\mathbf{Y}_\nu}
\newcommand{\yukt}[0]{\tilde{\mathbf{Y}}}

\newcommand{\ct}[0]{\dagger}
\newcommand{\tr}[0]{{\textsf{T}}}
\newcommand{\wt}[1]{\widetilde{#1}}

\newcommand{\GeV}[0]{\,\mathrm{GeV}}

\newcommand{\myArrayStretch}[0]{\renewcommand{\arraystretch}{1.3}}
\newcommand{\myArrayStretchReset}[0]{\renewcommand{\arraystretch}{1}}
\newcommand{\cM}[0]{\mathcal{M}}
\newcommand{\Vckm}[0]{V_\text{CKM}}
\newcommand{\Vpmns}[0]{V_\text{PMNS}}

\newcommand{\ah}[0]{{\hat{a}}}
\newcommand{\bh}[0]{{\hat{b}}}
\newcommand{\ch}[0]{{\hat{c}}}
\newcommand{\phant}[0]{{\phantom{\bh}}}

\DeclareMathOperator{\sign}{sign}
\DeclareMathOperator{\diag}{diag}

\def\BOX{\hbox to 3cm}

\def\MGUT{M_{\text{GUT}}}
\def\MSUSY{M_\text{SUSY}}
\def\MZ{M_{Z}}
\def\YQQ{\tilde{\mathbf{Y}}_{qq}}
\def\YEU{\tilde{\mathbf{Y}}_{eu}}
\def\YQL{\tilde{\mathbf{Y}}_{ql}}
\def\YUD{\tilde{\mathbf{Y}}_{ud}}

\def\MEFF{M^{\text{eff}}_{T}}

\begin{document}

\begin{titlepage}
\vspace*{0.7cm}

\begin{center}

	{\Large {\bf Employing nucleon decay as a fingerprint of SUSY GUT models \\[6pt] using \texttt{SusyTCProton}}}\\[8mm]

	Stefan Antusch$^{\star}$\footnote{Email: \texttt{stefan.antusch@unibas.ch}},  		Christian Hohl$^\star$\footnote{Email: \texttt{ch.hohl@unibas.ch}}, 
	and Vasja Susi\v{c}$^\star$\footnote{Email: \texttt{vasja.susic@unibas.ch}}

\end{center}

\vspace*{0.20cm}

\centerline{$^{\star}$ \it
Department of Physics, University of Basel,}
\centerline{\it
Klingelbergstr.\ 82, CH-4056 Basel, Switzerland}

\vspace*{1.2cm}

\begin{abstract}
\noindent 
While the observation of nucleon decay would be a smoking gun of Grand Unified Theories (GUTs) in general, the ratios between the decay rates of the various channels carry rich information about the specific GUT model realization. To investigate this fingerprint of GUT models in the context of supersymmetric (SUSY) GUTs, we present the software tool \texttt{SusyTCProton}, which is an extension of the module \texttt{SusyTC} to be used with the \texttt{REAP} package. It allows to calculate nucleon decay rates from the relevant dimension five GUT operators specified at the GUT scale, including the full loop-dressing at the SUSY scale. As an application, we investigate the fingerprints of two example GUT toy models with different flavor structures, performing an MCMC analysis to include the experimental uncertainties for the charged fermion masses and CKM mixing parameters. While both toy models provide equally good fits to the low energy data, we show how they could be distinguished via their predictions of ratios for nucleon decay rates. Together with \texttt{SusyTCProton} we also make the additional module \texttt{ProtonDecay} public. It can be used independently from \texttt{REAP} and allows to calculate nucleon decay rates from given $D=5$ and $D=6$ operator coefficients (accepting the required SUSY input for the $D=5$ case in SLHA format). The $D=6$ functionality can also be used to calculate nucleon decay in non-SUSY GUTs.
\end{abstract}

\end{titlepage}

\newpage
\tableofcontents
\newpage


\section{Introduction}

Grand Unified Theories (GUTs) \cite{Georgi:1974sy,Fritzsch:1974nn,Georgi:1974my} 
continue to present an interesting framework for physics Beyond the Standard Model (BSM). The attractive features of this class of theories are the unification of all gauge interactions at some high energy scale, (partial) unification of matter into representations of a bigger symmetry, and an explanation for the quantization of hypercharge in the Standard Model (SM). 

The presence and type of low energy phenomena (e.g.~light exotic matter) typically depends on the GUT model and as such these features are not universal predictions of gauge unification. An almost universal prediction of GUT models, however, is proton decay --- and more generally, nucleon decay. The gauge bosons transforming as $(\mathbf{3},\mathbf{2},-5/6)$ under the Standard Model gauge group $G_{321}\equiv \SU(3)\times\SU(2)\times\mathrm{U}(1)$ present in the adjoint representation of $\SU(5)$ violate baryon and lepton number ($B$ and $L$) symmetries, and mediate the process of nucleon decay, generating $B$ and $L$ violating four-fermion operators, i.e.~$D=6$ proton decay. Other particles, such as scalar leptoquarks, can of course also mediate proton decay, but the main point is that as long as the unified group contains $\SU(5)$ as a subgroup, proton (nucleon) decay will be present.

One advantage of nucleon decay as a probe of new physics is that our experiments are able to measure decay rates corresponding to very large masses of the mediators, and current experimental bounds demand that the GUT scale $\MGUT$ must be higher than around $10^{15}\GeV$. Since unification within field theory should happen below the Planck scale, this implies a window between roughly $10^{15}\GeV$--$10^{19}\GeV$ for unification. The renormalization group (RG) running of gauge couplings in the SM can be made consistent with this window in the presence of appropriate intermediate states.

In supersymmetric (SUSY) theories, assuming the effective theory above the SUSY scale $\MSUSY$ to be the Minimal Supersymmetric Standard Model (MSSM), it is possible to construct $B$ and $L$ violating superpotential operators of four fields of the type $qqql$, where $q$ stands for a quark and $l$ for a lepton field. Such operators are generated, for example, when integrating out chiral supermultiplets of heavy color triplets $(\mathbf{3},\mathbf{1},-1/3)$ under $G_{321}$. The resulting Lagrangian operators are dimension $5$ operators, consisting of $2$ fermions and $2$ scalar superpartners of SM fermions. The $D=5$ diagrams are then dressed at the SUSY scale to have SM external states, thus generating four-fermion nucleon decay operators. Contributions to nucleon decay generated in SUSY in such a way are referred to as $D=5$ proton decay \cite{Sakai:1981pk,Weinberg:1981wj}, and they typically dominate over $D=6$ contributions, also usually requiring a higher GUT scale $\MGUT\gtrsim 10^{16}\GeV$ in SUSY GUTs compared to non-SUSY GUTs (due to the requirement of sufficiently large triplet masses). 

A rough estimate of the total proton decay rate can be obtained using the formula
\begin{align}
\Gamma &\approx |\mathcal{A}|^{2}\,m_p^5,\label{eq:estimate-proton}
\end{align}
where $m_p\approx 0.94\GeV$ is the proton (nucleon) mass and the $D=5$ amplitude $A_{D=5}$ for the process can be estimated, see e.g.~\cite{Senjanovic:2009kr}, as
\begin{align}
\mathcal{A}_{D=5}&\approx \frac{\alpha}{4\pi}\,y_u y_d\;\frac{m_{\tilde{w}}}{m_{T}m_{\tilde{f}}^{2}},\label{eq:estimate-proton5}
\end{align}
with $\alpha=g^2/(4\pi)$, $g$ the unified gauge coupling, $y_u$ and $y_d$ the Yukawa couplings of the up- and down-quarks, $m_{\tilde{w}}$ the gaugino mass, $m_{\tilde{f}}$ the sfermion mass, and $m_{T}$ the mass of the heavy color triplet mediating the decay. We assumed that the $D=6$ contribution, e.g.~from gauge bosons, is negligible, i.e.~$|\mathcal{A}_{D=5}|\gg |\mathcal{A}_{D=6}|$, since its amplitude is proportional to the square of the inverse heavy mass:
\begin{align}
\mathcal{A}_{D=6}&\approx \frac{4\pi\alpha}{m_{X}^{2}},\label{eq:estimate-proton6}
\end{align}
with $m_{X}$ the mass of the gauge bosons mediating the $B$ and $L$ violating interaction. A thorough analysis of nucleon decay, however, should determine the decay rates into each channel separately, and both $D=5$ and $D=6$ contributions should be taken into account (we assume $R$-parity conservation, so no $D=4$ proton decay).

It is well known, see e.g.~\cite{Nath:2006ut}, that in SUSY theories the dominant decay channel is usually the kaon-neutrino channel $p\to K^{+}\bar{\nu}$. The proton lifetime bound for this channel currently comes from Super-Kamiokande~\cite{Mine:2016mxy}: $6.6\cdot 10^{33}$ years at $90\,\%$~C.L. Future experiments are set to improve on this bound further: the expected $90\,\%$~C.L.~bounds after 10 years of operation are $3.3\cdot 10^{34}$ years for Dune~\cite{Acciarri:2015uup}, $3.0\cdot 10^{34}$ years for Hyper-Kamiokande~\cite{Abe:2018uyc} and $2.0\cdot 10^{34}$ years for JUNO~\cite{An:2015jdp}.

To investigate the viability of a specific SUSY GUT model and determine its predictions, at least as far as nucleon decay is concerned, a more precise estimate of proton/nucleon decay rates is necessary than that of Eq.~\eqref{eq:estimate-proton5}. While the procedure and expressions for the decay rates of the various nucleon decay channels are known, cf.~\cite{Goto:1998qg,Nath:2006ut}, they are cumbersome to implement from scratch.
A software package capable of the computation of nucleon decay rates in the various decay channels would thus be most welcome. Ideally, this tool would be able to handle all the complications in the computation that can arise, in particular both the sparticle spectrum and the flavor violating effects in the SUSY sector can play a prominent role. Furthermore, the effective $D=5$ operators arising from integrating out the heavy triplets may also carry a flavor imprint from the Yukawa sector, since both the couplings to the heavy triplets and MSSM Higgs doublets arise from the same GUT operators. To be able to handle all of this automatically, the tool would thus ideally be built on top of an existing 
SUSY spectrum generator of sufficient sophistication.   

In this paper we roll out exactly such a nucleon decay calculation tool for SUSY GUTs consisting of two in principle independent Mathematica modules \texttt{SusyTCProton.m} and \texttt{ProtonDecay.m} that we make publicly available:\footnote{A User's Guide for the installation and use of these packages is provided in Appendix~\ref{sec:user_guide}.}
\begin{enumerate}
\item 
The \texttt{SusyTCProton} package is a nucleon decay extension of the existing \texttt{SusyTC} package~\cite{Antusch:2015nwi} built within the framework of \texttt{REAP}~\cite{Antusch:2005gp}. The standard \texttt{SusyTC} performs the RG evolution of the softly broken MSSM from the GUT scale to the $Z$-boson scale, automatically matching the MSSM and SM at an appropriate scale and generating all the SUSY sector information; the nucleon decay extension of this package performs, in addition, the $1$-loop running of $D=5$ operators from the GUT to the SUSY scale.  
\item 
The \texttt{ProtonDecay} package takes as input the $D=5$ decay operators and the sparticle masses and mixings, all at the SUSY scale, and dresses the relevant diagrams with SUSY sparticles to determine the effective four-fermion operators of nucleon decay, and after RG running to the nucleon scale computes the decay rates in $8$ proton decay channels and $5$ neutron decay channels (not counting different neutrino flavors, see Table~\ref{tab:decay-lifetimes} for the list). The input for this package is compatible both with a \texttt{SusyTCProton} output of internal values, as well as with an SLHA file~\cite{Skands:2003cj} amended with the $D=5$ operator information. This allows \texttt{ProtonDecay} to be used also with any spectrum generator that can output an SLHA file, such as SPheno~\cite{Porod:2003um,Porod:2011nf}, SOFTSUSY~\cite{Allanach:2001kg}, SuSeFLAV~\cite{Chowdhury:2011zr} or SuSpect~\cite{Djouadi:2002ze}, provided that the user inputs $D=5$ operator values from their own source. In addition, the \texttt{ProtonDecay} package can handle $D=6$ contributions to nucleon decay, either as standalone terms (the non-SUSY GUT case) or in tandem with $D=5$ contributions; the user needs to input them at a scale below which only the SM RG running is applied.
\end{enumerate}
As a demonstration, we employ the above software packages in this paper to analyze the nucleon decay in two example toy models, cf.~\cite{Antusch:2014poa,Antusch:2013kna}, both of which are of the flavor SUSY GUT variety. These models fit the SM fermion masses and mixings through particular choices of Yukawa textures and GUT-operators, which generate the Yukawa sector. The type of GUT operators used implies a choice of the Clebsch-Gordan (CG) coefficients between different sectors associated by the GUT symmetry. These same features impact also the $D=5$ proton decay predictions, and it is precisely these flavor effects on nucleon decay that we focus on here. The magnitude of the decay rates in the various channels is subject to various potential ambiguities, such as the mass of the heavy triplet. These are overcome, at least in the models we consider here, by investigating ratios of decay rates, allowing to identify a nucleon decay ``fingerprint'' for each model. If nucleon decay events are seen experimentally in at least one decay channel, such a fingerprint technique, as we demonstrate in this paper, can be a powerful tool to either discriminate within a set of models with equally good Yukawa-sector fits and near-identical SUSY spectra, or to possibly rule out particular models of this type.   

The paper is organized as follows. We define the example flavor GUT models in Section~\ref{sec:model_setup}, discuss how we implemented our numerical analysis in Section~\ref{sec:procedure-for-numerical-analysis}, present our results regarding nuclear decay in these models in Section~\ref{sec:results} and give our conclusions in Section~\ref{sec:conclusions}. The paper contains also 3 appendices. Appendix~\ref{sec:DT-splitting} presents a discussion on our doublet-triplet splitting setup and a few of its concrete realizations. Appendix~\ref{sec:documentation} provides technical details on the nucleon decay computation and a note on conventions. Finally, Appendix~\ref{sec:user_guide} contains the documentation for our software packages \texttt{SusyTCProton} and \texttt{ProtonDecay}.


\section{Choice of two example models for investigating nucleon decay \label{sec:model_setup}}
\subsection{General setup in flavor SUSY GUTs\label{sec:general-setup}}

The models under consideration are flavor $\mathrm{SU}(5)$ SUSY GUTs of the type discussed in~\cite{Antusch:2014poa,Antusch:2013kna}. In this paper, we consider $2$ toy models of this type and analyze their nucleon decay predictions.

The 3 generations of SM fermions are embedded in the standard way into $\SU(5)$ representations, where the index $i$ runs from $1$ to $3$:

\begin{align}
\mathbf{10}_{Fi}&=Q_i\oplus u^{c}_{i}\oplus e^{c}_{i},\\
\mathbf{\bar{5}}_{Fi}&=L_{i}\oplus d^{c}_i.
\end{align}

\noindent
The superpotential of the models can be separated into the Yukawa and breaking part
\begin{align}
W&=W_{\text{Yuk}}+W_{\text{breaking}},
\end{align}

\noindent
where the breaking part consists of those terms, by definition, which do not contain the ``fermionic'' chiral representations $\mathbf{10}_{Fi}$ and $\mathbf{\bar{5}}_{Fi}$. For each model we assume a particular texture of Yukawa matrices, to be specified later, and \textit{single operator dominance} in each Yukawa entry, i.e.~only a single GUT operator is present or dominates every Yukawa entry (see \cite{Antusch:2019avd,Antusch:2018gnu,Antusch:2017ano,Antusch:2013kna,Antusch:2012fb} for concrete realizations in the literature). This imposes strict relations between different fermion sectors and increases model predictivity. The Yukawa sector of each model thus consists of a set of (non-renormalizable) operators.

We omit the high-energy model building in the neutrino sector. Also, we remain agnostic about the details of the breaking sector; the only requirement is that it breaks $\SU(5)$ symmetry to the Standard Model group $G_{321}:=\SU(3)\times\SU(2)\times\mathrm{U}(1)$ and includes the Higgs chiral supermultiplets assumed to be present due to their use in the Yukawa sector.

We are interested in models with as predictive $D=5$ proton decay as possible. Since this mode of proton decay is mediated by color triplets $(\mathbf{3},\mathbf{1},-1/3)$ of $G_{321}$, the details of doublet-triplet (DT) splitting come into the decay prediction. We thus consider 
as simple a scenario as possible for DT splitting, where only one triplet pair $T_1\oplus\overline{T}_1$ couples to the fermionic sector, and for simplicity we assume that it is coming from $\mathbf{5}_H\oplus \mathbf{\bar{5}}_H$ of $\SU(5)$.

This puts a limitation on which flavor GUT models can be considered in the context of this paper: the Higgs cannot be contained in the $\mathbf{45}$ or $\mathbf{\overline{45}}$ of $\SU(5)$, and the non-renormalizable operators making use of such representations in the lists provided in \cite{Antusch:2009gu,Antusch:2013rxa} are not available, but only those containing $\mathbf{5}_H$ or $\mathbf{\bar{5}}_H$.

At the GUT scale $\MGUT$, the GUT models are matched to the softly broken MSSM as the effective theory, with the addition of the effective $D=5$ proton decay operators in the superpotential. For the soft terms, we assume constrained MSSM (cMSSM, a.k.a.~mSUGRA) boundary conditions at $\MGUT$ (see e.g.~\cite{Martin:1997ns}), so they are parametrized by 
\begin{align}
m_0, \quad M_{1/2}, \quad A_0,
\end{align}
where $m_0$ is the universal soft scalar mass, $M_{1/2}$ the universal gaugino mass and $A_{0}$ the universal $A$-term factor. The choice of cMSSM is in principle not necessary, but it ensures that our analysis isolates the effects on nucleon decay from the flavor structure of the theory without added flavor ambiguities from the SUSY sector.

At the level of the MSSM, the Yukawa sector (including the relevant operators for proton decay, but ignoring any neutrino terms) consists of

\begin{align}
W_{\text{Yuk}}&=\sum_{i,j=1}^{3}\;\; (\YU)_{ij}\;Q_i\cdot H_u\,u_j^c-
(\YD)_{ij}\;Q_i\cdot H_d\,d^c_j-(\YE)_{ij}\;L_i\cdot H_d\,e^c_j+W_T,
\label{eq:MSSM-superpotential}
\end{align}
where the color indices have been suppressed and the $\SU(2)$ contraction is written as $\Psi\cdot \Phi=\epsilon_{kl} \Psi^k \Phi^l$, with $\epsilon_{kl}$ the $2$-index completely anti-symmetric tensor in the convention $\epsilon_{12}=+1$. 
Notice that we have written the Yukawa terms in the left-right (LR) convention. 

The triplet part of the Yukawa superpotential consists of 
\begin{align}
\begin{split}
W_\text{T} &= + \tfrac{1}{2}(\YQQ)_{ij}\,\epsilon_{\ah\bh\ch}\,Q_i^\ah\cdot Q_j^\bh\, T_1^\ch + (\YEU)_{ij}\,e^c_i\,u^c_{j\ah}\,T_1^\ah \\
&\quad + (\YQL)_{ij}\,Q_i^\ah\cdot L_j\,\overline{T}_{1\ah} + (\YUD)_{ij}\,\epsilon^{\ah\bh\ch}\,u^c_{i\ah}\,d^c_{j\bh}\,\overline{T}_{1\ch} \\
&\quad + \sum_{I,J}(\mathbf{M}_T)_{IJ}\,T_I^\ah\,\overline{T}_{J\ah},
\end{split} 
\end{align}
where $\ah,\bh,\ch$ are $\SU(3)$ indices, while $\epsilon_{\ah\bh\ch}$ and 
$\epsilon^{\ah\bh\ch}$ are Levi-Civita tensors with $\epsilon_{123}=\epsilon^{123}=1$. The $\SU(2)$ contractions are again labeled with a $\cdot$ dot. We refer to the $3\times 3$ matrices $\YQQ$, $\YEU$, $\YQL$ and $\YUD$ as \textbf{quasi-Yukawa} matrices in the remainder of this text.

This ansatz is specific to the models under consideration, since only the (anti-)triplets of $\mathbf{5}_H\oplus\mathbf{\bar{5}}_H$ couple to the SM fermions; we thus omit any triplet indices $I,J$ in the quasi-Yukawas. The general case with an arbitrary number of triplets is considered in Appendix~\ref{appendix:quasiYukawa-definitions}; the quasi-Yukawas written above would have $I,J=1$, while all other quasi-Yukawas for $I,J\neq 1$ would be zero. 

Integrating out the triplets and adapting the result from Appendix~\ref{appendix:quasiYukawa-definitions} to our specific case, we obtain the following effective $D=5$ proton decay operators:
\begin{align}
W_T&\to -\tfrac{1}{2}\mathbf{C}^{ijkl}_{5L}\,\epsilon_{\ah\bh\ch}\,(Q_i^\ah\cdot L_j)(Q_k^\bh\cdot Q_l^\ch) - \mathbf{C}^{ijkl}_{5R}\,\epsilon^{\ah\bh\ch}\,u^c_{i\ah}\,d^c_{j\bh}\,e^c_k\,u^c_{l\ch}\,,
\end{align}
where the $\mathbf{C}_{5}$ coefficients for our models equal
\begin{align}
\mathbf{C}^{ijkl}_{5L} &= (1/\MEFF)\,(\YQL)_{ij}\,(\YQQ)_{kl},\label{eq:C5-in-model-begin}\\
\mathbf{C}^{ijkl}_{5R} &= (1/\MEFF)\,(\YUD)_{ij}\,(\YEU)_{kl},\label{eq:C5-in-model-end}
\end{align}
with 
\begin{align}
\MEFF:=((\mathbf{M}_T^{-1})_{11})^{-1} \label{eq:effective-triplet-definition}
\end{align}
denoting the ``triplet effective mass''.  We provide a more detailed discussion on the form of the $2\times 2$ matrix $\mathbf{M}_T$
and the effective triplet mass in Appendix~\ref{sec:DT-splitting}. It is clear, however, that in this scheme the effective triplet mass does not necessarily correspond to any physical scale of particles, and as such can have a value above the Planck scale without losing consistency.

In the $\SU(5)$ context, the Yukawa and quasi-Yukawa matrices come from the following GUT operators:
\begin{align}
\mathbf{10}_F\,\mathbf{10}_F\,\mathbf{X}&\quad\supset\quad \YU,\YQQ,\YEU,\\
\mathbf{10}_F\,\mathbf{\bar{5}}_F\,\mathbf{X}&\quad\supset\quad \YD,\YE,\YQL,\YUD,
\end{align}
where $\mathbf{X}$ is a placeholder for one or multiple Higgs fields in a Yukawa operator, possibly different for different matrix entries. The matrices arising from the same operator are related via $\SU(5)$ CG coefficients, which depend on the combination of Higgs fields $\mathbf{X}$ (entry by entry).

\subsection{Specific models: their Yukawa and quasi-Yukawa sectors}

We consider two models in this paper, referred to as model 1 and model 2. They in fact make use of the Yukawa textures of the models constructed in section 4 of \cite{Antusch:2014poa}. We are interested only in the Yukawa sector here, so we remain agnostic about the details of doublet-triplet splitting (see Appendix~\ref{sec:DT-splitting} for some possibilities).

The Yukawa textures of the two models are:
\begin{align}
\text{model 1}:&&
\mathbf{Y}_{\mathbf{10}}&=\begin{pmatrix}\ast &\ast&0\\\ast&\ast&\ast\\0&\ast&\ast\\\end{pmatrix},
&
\mathbf{Y}_{\mathbf{\bar{5}}}&=\begin{pmatrix}0&\ast&0\\\ast&\ast&0\\0&0&\ast\\\end{pmatrix},\label{eq:model1-Yukawa-texture}\\
\text{model 2}:&&
\mathbf{Y}_{\mathbf{10}}&=\begin{pmatrix}\ast&\ast&\ast\\\ast&\ast&\ast\\\ast&\ast&\ast\\\end{pmatrix},
&
\mathbf{Y}_{\mathbf{\bar{5}}}&=\begin{pmatrix}\ast&0&0\\0&\ast&0\\0&0&\ast\\\end{pmatrix},\label{eq:model2-Yukawa-texture}
\end{align}
where $\ast$ denotes a non-vanishing complex entry and $\mathbf{Y}_{\mathbf{10}}$ are always symmetric: $\mathbf{Y}_{\mathbf{10}}=\mathbf{Y}_{\mathbf{10}}^\tr$. Note that in order to consider the two models on the same footing, we do not assume any spontaneous CP violation mechanism\footnote{The use of a spontaneous CP violation mechanism \cite{Antusch:2011sx,Antusch:2013wn} in model 1 could force the $(\mathbf{Y}_{\mathbf{10}})_{12}$ entry imaginary and the other Yuakwa parameters real \cite{Antusch:2009hq,Antusch:2018gnu}; this would automatically give a correct prediction for the CKM CP-violating phase.} of model 1, so the parameters are taken as complex.

The non-zero entries in these matrices are populated by the operators specified in Table~\ref{tab:Yukawa-entry-operators-for-models} (chosen from Table~1 of \cite{Antusch:2009gu}). The $\SU(5)$ representation in the index position indicates the type of contraction used between the pieces (analogous to integrating out the mediator in the index from a renormalizable interaction of that mediator with the fields in the parentheses). The vacuum expectation value (VEV) of the $\mathbf{24}_H$ is at the GUT scale.

\def\TABSKIP{8pt}

\newcommand{\OPFORTYFIVE}[2]{(\mathbf{10}_{F{#1}})_{\mathbf{\overline{45}\otimes\mathbf{\overline{45}}}}\;(\mathbf{\bar{5}}_{H}\mathbf{24}_H)_{\mathbf{45}}\;(\mathbf{\bar{5}}_{F{#2}}\mathbf{24}_H)_{\mathbf{45}}}

\newcommand{\OPTEN}[2]{(\mathbf{10}_{F#1}\mathbf{24}_H)_{\mathbf{\overline{10}}}\;(\mathbf{\bar{5}}_{F#2
}\mathbf{\bar{5}}_H)_{\mathbf{10}}}

\newcommand{\OPFIVE}[2]{(\mathbf{10}_{F{#1}}\mathbf{\bar{5}}_H)_{\mathbf{\bar{5}}}\;(\mathbf{10}_{F{#2}}\mathbf{24}_H)_{\mathbf{5}}}

\begin{table}[htb]
\caption{Table of operators used in each Yukawa entry for models 1 and 2.\label{tab:Yukawa-entry-operators-for-models}}
\begin{center}
\vskip -0.6cm
\begin{tabular}{ll@{\qquad}l}
\toprule
entry&model 1 operator&model 2 operator\\
\midrule
$(\mathbf{Y}_{\mathbf{10}})_{ij}$&
$\mathbf{10}_{Fi}\mathbf{10}_{Fj}\mathbf{5}_{H}$&
$\mathbf{10}_{Fi}\mathbf{10}_{Fj}\mathbf{5}_{H}$\\[\TABSKIP]
$(\mathbf{Y}_{\mathbf{\bar{5}}})_{11}$&
/&
$\OPFORTYFIVE{1}{1}$\\[\TABSKIP]
$(\mathbf{Y}_{\mathbf{\bar{5}}})_{12}$&
$\OPFORTYFIVE{1}{2}$&
/\\[\TABSKIP]
$(\mathbf{Y}_{\mathbf{\bar{5}}})_{21}$&
$\OPTEN{2}{1}$&
/\\[\TABSKIP]
$(\mathbf{Y}_{\mathbf{\bar{5}}})_{22}$&
$\OPTEN{2}{2}$&
$\OPTEN{2}{2}$\\[\TABSKIP]
$(\mathbf{Y}_{\mathbf{\bar{5}}})_{33}$&
$\OPFIVE{3}{3}$&
$\OPFIVE{3}{3}$\\
\bottomrule
\end{tabular}
\end{center}
\end{table}

This set of operators gives the following CG factor relations for the Yukawas and quasi-Yukawas, which we computed explicitly (the reader can also cross-check with \cite{Antusch:2009gu}):
\begin{itemize}
\item model 1:
\begin{align}
\YE &= \begin{pmatrix}
\phantom{+}0 & \phantom{+}6 & \phantom{+}0 \\
-\tfrac{1}{2} & \phantom{+}6 & \phantom{+}0 \\
\phantom{+}0 & \phantom{+}0 & -\tfrac{3}{2}
\end{pmatrix} \cdot \YD^\tr\,,
&\YQL &= \begin{pmatrix}
\phantom{+}0 & -1 & \phantom{+}0 \\
-1 & -1 & \phantom{+}0 \\
\phantom{+}0 & \phantom{+}0 & \phantom{+}\tfrac{3}{2}
\end{pmatrix} \cdot \YD\,,
&\YUD &= \begin{pmatrix}
\phantom{+}0 & \phantom{+}\tfrac{2}{3} & \phantom{+}0 \\
-4 & -4 & \phantom{+}0 \\
\phantom{+}0 & \phantom{+}0 & \phantom{+}1
\end{pmatrix} \cdot \YD\,,\label{eq:model1-Clebsch-begin}\\
\yukt_{qq} &= - \YU\,, &
\yukt_{eu} &= + \YU\,.\label{eq:model1-Clebsch-end}
\end{align}
\item model 2:
\begin{align}
\YE &= \begin{pmatrix}
-\tfrac{1}{2} & \phantom{+}0 & \phantom{+}0 \\
\phantom{+}0 & \phantom{+}6 & \phantom{+}0 \\
\phantom{+}0 & \phantom{+}0 & -\tfrac{3}{2}
\end{pmatrix} \cdot \YD^\tr\,,
&\YQL &= \begin{pmatrix}
-1 & \phantom{+}0 & \phantom{+}0 \\
\phantom{+}0 & -1 & \phantom{+}0 \\
\phantom{+}0 & \phantom{+}0 & \phantom{+}\tfrac{3}{2}
\end{pmatrix} \cdot \YD\,,
&\YUD &= \begin{pmatrix}
\phantom{+}\tfrac{2}{3} & \phantom{+}0 & \phantom{+}0 \\
\phantom{+}0 & -4 & \phantom{+}0 \\
\phantom{+}0 & \phantom{+}0 & \phantom{+}1
\end{pmatrix} \cdot \YD\,,\label{eq:model2-Clebsch-begin}\\
\yukt_{qq} &= - \YU\,, &
\yukt_{eu} &= + \YU\,.\label{eq:model2-Clebsch-end}
\end{align}
\end{itemize}
The dot $\cdot$ indicates entry-wise multiplication of matrices.

\section{Considerations and procedure for a numerical analysis \label{sec:procedure-for-numerical-analysis}}

\subsection{Implementation of the Yukawa sector \label{sec:implementation}}
Following the considerations in Section~\ref{sec:model_setup}, we implement models~1 and 2  at the GUT scale in the framework of the MSSM. The Yukawa sector is determined by specifying the two GUT Yukawa matrices $\mathbf{Y}_{\mathbf{\bar{5}}}$ and $\mathbf{Y}_{\mathbf{10}}$, which is equivalent to specifying the up- and down-quark matrices $\mathbf{Y}_{u}$ and $\mathbf{Y}_d$, since all the other Yukawa and quasi-Yukawa parameters are related to them via CG factors, see Eq.~\eqref{eq:model1-Clebsch-begin}-\eqref{eq:model1-Clebsch-end} and \eqref{eq:model2-Clebsch-begin}-\eqref{eq:model2-Clebsch-end} for model 1 and 2, respectively.

Given the textures of the two models, $\mathbf{Y}_{u}$ is a complex symmetric matrix, with the added constraint $(\mathbf{Y}_{u})_{13}=(\mathbf{Y}_{u})_{31}=0$  on its entries in the case of model 1. Any symmetric complex $3\times 3$ matrix $\mathbf{Y}$ can be decomposed, using Takagi decomposition, as

\begin{align}
\mathbf{Y}=\mathbf{U}\;\diag(y_1,y_2,y_3)\;\mathbf{U}^\tr,\label{eq:Takagi-decomposition}
\end{align}
where the singular values $y_i\geq 0$, and $\mathbf{U}$ is a $3\times 3$ unitary matrix. For the parametrization of $\mathbf{U}$, we shall make use of the definitions
\begin{align}
\mathbf{U}_{12}(\theta,\delta) &:= \begin{pmatrix} \phantom{-}\cos\theta\,\phantom{e^{i\delta}} & \sin\theta\,e^{-i\delta} & 0 \\ -\sin\theta\,e^{i\delta}& \cos\theta & 0 \\ 0 & 0 & 1 \end{pmatrix},\\[6pt]
\mathbf{U}_{23}(\theta,\delta) &:= \begin{pmatrix} 1 & 0 & 0 \\ 0 & \phantom{+}\cos\theta\,\phantom{e^{i\delta}}& \sin\theta\,e^{-i\delta} \\ 0 & -\sin\theta\,e^{i\delta} & \cos\theta \end{pmatrix}, \\[6pt]
\mathbf{U}_{13}(\theta,\delta) &:= \begin{pmatrix} \phantom{-}\cos\theta\,\phantom{e^{i\delta}} & 0&\sin\theta\,e^{-i\delta} \\ 0&1&0\\ 
-\sin\theta\,e^{i\delta}& 0&\cos\theta\\ \end{pmatrix},
\end{align}
where $\mathbf{U}_{12}$, $\mathbf{U}_{23}$ and $\mathbf{U}_{13}$ represent unitary matrices of $2$-D complex rotations, each in principle depending upon an angle and a phase. The most general $3\times 3$ unitary matrix can be explicitly written by using the following family of real parameterizations:
\begin{align}
\mathbf{U}&= \diag(e^{i\eta_{1}},e^{i\eta_{2}},e^{i\eta_{3}})
\;\mathbf{U}_{23}(\theta^{uL}_{23},\delta_{0})
\;\mathbf{U}_{13}(\theta^{uL}_{13},\delta) 
\;\mathbf{U}_{12}(\theta^{uL}_{12},0)
\;\diag(e^{i\phi_{1}/2},e^{i\phi_{2}/2},1),\label{eq:parametrization-unitary-3by3}
\end{align} 

\noindent
where the family parameter $\delta_{0}$ is fixed, and the $9$ free parameters describing the unitary rotation consist of angles $\theta_{ij}$ and phases $\delta$, $\phi_i$ and $\eta_i$: 
\begin{align}
\theta_{12},
\quad \theta_{13},
\quad \theta_{23},
\quad \delta,
\quad \phi_1,
\quad \phi_2,
\quad \eta_{1},
\quad \eta_{2},
\quad \eta_{3}.
\end{align}

The first and last factor of the product in Eq.~\eqref{eq:parametrization-unitary-3by3} are the two diagonal matrices containing the ``outer phases'' $\eta_{i}$ and the ``inner phases'' $\phi_i$ (with inner/outer referring to their location in the product of Eq.~\ref{eq:Takagi-decomposition}). There are only $2$ $\phi$-phases, since a third such phase can be absorbed as a common phase factor into the $\eta$ phases. Furthermore, the $1/2$ factors for the $\phi$-phases are there for later convenience; we shall view them as phases of the singular values in the Takagi decomposition rather than as part of the unitary rotation.

Due to greater convenience, we use different parameterizations in the two models. In model~1, we set $\delta_{0}=\pi/2$ and impose the additional (complex) condition $(\mathbf{Y}_{u})_{13}=0$ by solving for the real variables $\theta_{13}$ and $\delta$. For model~2 we set $\delta_{0}=0$. The advantages of this choice of $\delta_{0}$ will be discussed later for each model separately.

The use of Takagi decomposition for the parametrization of $\YU$ has the advantage that it provides direct control over the singular values (Yukawa couplings) and mixing angles in the up-quark Yukawa matrix. Also, note that we can redefine each fermionic representation $\mathbf{10}_{Fi}$ and $\bar{\mathbf{5}}_{Fi}$ with an arbitrary complex phase, implying that $6$ phases in the Yukawa sector are unphysical. In our convenient parameterization of $\mathbf{Y}_u$, a phase redefinition of $\mathbf{10}_{Fi}$ allows us to fix the outer phases without loss of generality: 
\begin{align}
\eta_{1}=\eta_{2}=\eta_{3}&=0.
\end{align}

We shall use the phase redefinitions of $\bar{\mathbf{5}}_{Fi}$ to remove the phase of one complex entry in each column of $\mathbf{Y}_d$. In model~2, this implies that
$\mathbf{Y}_d$ is diagonal with (positive) real values, while in model~1 
a complex phase remains in one $2$nd column entry; we choose that entry to be
$(\mathbf{Y}_d)_{12}$.

Putting all the above considerations together, the final form of the parameterizations of the Yukawa matrices $\mathbf{Y}_u$ and $\mathbf{Y}_d$, which uniquely determine the entire Yukawa sector, are the following:

\begin{itemize}
\item \textbf{Model~1:}\\
According to the textures in Eq.~\eqref{eq:model1-Yukawa-texture}, the parametrization of the down- and up-quark Yukawa matrix in the context of the MSSM at the GUT scale is given by
\begin{align}
\YD&=\begin{pmatrix}
0 & y^d_{12}\,e^{i\varphi} & 0 \\
y^d_{21} & y^d_{22} & 0 \\
0 & 0 & y^d_{3}
\end{pmatrix}, \label{eq:model1-implementation-yd}\\[6pt]
\YU&= \mathbf{U}_1\; \text{diag}(y^u_{1}\, e^{i\phi_1},y^u_{2}\, e^{i\phi_2},y^u_{3})\;\mathbf{U}_1^\tr, \label{eq:model1-implementation-yu-1}
\end{align}
with
\begin{align}
\mathbf{U}_1&=\mathbf{U}_{23}(\theta^{uL}_{23},\pi/2)
\; \mathbf{U}_{13}(\theta^{uL}_{13},\delta)
\; \mathbf{U}_{12}(\theta^{uL}_{12},0), \label{eq:model1-implementation-yu-2}
\end{align}
where the values of $\theta^{uL}_{13}$ and $\delta$ depend on the other parameters of the up-sector; they are determined from demanding $(\mathbf{Y}_u)_{13}=0$:\footnote{We omit an overall $\cos\theta^{uL}_{13}$ factor in the $(\mathbf{Y}_u)_{13}=0$ condition. Such a factor allows for an alternative solution $\theta^{uL}_{13}=\pi/2$, which also sets $(\mathbf{Y}_u)_{23}=0$, so we reject it.}
\begin{align}
(C_1 e^{i\delta}+C_{2} e^{-i\delta})\sin\theta^{uL}_{13}+C_{3}&=0,\label{eq:model1-C-constraint}
\end{align}
with
\begin{align}
\begin{split}
C_{1}&:= y^{u}_{3}\,\cos\theta^{uL}_{23},\\
C_{2}&:= (y^{u}_{2}e^{i\phi_2}\sin^{2}\theta^{uL}_{12}-y^{u}_{1}e^{i\phi_1}\cos^{2}\theta^{uL}_{12})\,\cos\theta^{uL}_{23},\\
C_{3}&:=i\,(y^{u}_{1}e^{i\phi_1}+y^{u}_{2}e^{i\phi_2})\,\sin\theta^{uL}_{12}\,\cos\theta^{uL}_{12}\,\sin\theta^{uL}_{23}.\\ 
\end{split}\label{eq:model1-C-coefficients}
\end{align}
Once the numeric values of the free parameters are specified, the $C$ coefficients in Eq.~\eqref{eq:model1-C-coefficients} become complex numbers, allowing for the complex equation~\eqref{eq:model1-C-constraint} to be solved numerically. Given the hierarchical structure of masses and mixings in the quark sector, this equation can always be solved for $\delta$ and $\theta^{uL}_{13}$ in a numerically stable fashion. The values for the angle are very small: $\theta^{13}\lesssim 10^{-5}$, which suggests a possible alternative which we mention here in passing: we could omit solving Eq.~\eqref{eq:model1-C-constraint} and simply take $\theta^{uL}_{13}=0$ ($\delta$ dependence then disappears). This would be a relatively good approximation yielding $|(\YU)_{13}|\lesssim 10^{-5}\approx 0$.

The CKM matrix comes from diagonalizing both $\YU$ and $\YD$. It turns out that for this particular texture, choosing the $\delta_{0}=\pi/2$ parametrization for the up-sector then already provides a good CKM phase for $\varphi\approx 0$ in $\YD$. 

Also, note that the complex phase $\phi$ here is in the $(\YD)_{12}$ entry, in contrast to the class of models in~\cite{Antusch:2018gnu} where a phase is in the $(\YD)_{21}$ entry. Despite the same texture, the difference comes from removing the phases in $\YD$ 
using $\bar{\mathbf{5}}_{Fi}$ in our case (removes one phase per column), while \cite{Antusch:2018gnu} prefers to remove phases via $\mathbf{10}_{Fi}$ redefinitions (removes one phase per row), so that they can make use of the $\bar{\mathbf{5}}_{Fi}$ freedom in their implementation of the neutrino sector, which we do not consider in the present model.

The charged lepton Yukawa matrix $\YE$ and the quasi Yukawa matrices $\YQQ$, $\YEU$, $\YQL$ and $\YUD$ for model~1 are determined from Eq.~\eqref{eq:model1-Clebsch-begin} and \eqref{eq:model1-Clebsch-end}.

\item \textbf{Model~2:}\\
Corresponding to the textures in Eq.~\eqref{eq:model2-Yukawa-texture}, the parametrization of the down- and up-quark Yukawa matrix in the context of the MSSM at the GUT scale is given by
\begin{align}
\YD&= \text{diag}(y^{d}_{1},y^{d}_{2},y^{d}_{3}),\label{eq:model2-implementation-yd}\\
\YU&= \mathbf{U}_2\; \text{diag}(y^u_{1}\, e^{i\phi_1},y^u_{2}\, e^{i\phi_2},y^u_{3})\;\mathbf{U}_2^\tr,\label{eq:model2-implementation-yu-1}
\end{align}
where
\begin{align}
\mathbf{U}_2&=\mathbf{U}_{23}(\theta^{uL}_{23},0)\;\mathbf{U}_{13}(\theta^{uL}_{13},\delta)\; \mathbf{U}_{12}(\theta^{uL}_{12},0),\label{eq:model2-implementation-yu-2}
\end{align}
while $\YE$, $\YQQ$, $\YEU$, $\YQL$ and $\YUD$ are then determined from Eq.~\eqref{eq:model2-Clebsch-begin} and \eqref{eq:model2-Clebsch-end}. 

In model~2, the CKM matrix is controlled entirely by the up-sector: choosing the 
$\delta_{0}=0$ parametrization in the up-sector thus provides the standard parametrization of the CKM matrix and the $3$ angles and phases in $\mathbf{U}_2$ are directly those of the CKM values at the GUT scale.
\end{itemize}

\subsection{Model parameters \label{sec:parameters}}
Given the implementation/parameterization of the Yukawa sectors of the two models in the previous subsection, we can now turn our attention to comprehensively analyzing the parameters and observables. The two models contain the following input parameters at the GUT scale:
\begin{itemize}

\item \textbf{Model~1} contains the input parameters
\begin{align}
y^d_{12},\ y^d_{21},\ y^d_{22},\ y^d_{3},\ y^u_{1},\ y^u_{2},\ y^u_{3},\ \theta^{uL}_{12},\ \theta^{uL}_{23},\ \varphi,\ \phi_1,\ \phi_2;\quad m_0,\ M_{1/2},\ A_{0},\ \tan\beta,\label{eq:model1-input-parameters}
\end{align}
where the Yukawa sector parameters $y^{u,d}$, $\theta^{uL}$, $\varphi$ and $\phi$ are used for the constructions in Eqs.~\eqref{eq:model1-implementation-yd}--\eqref{eq:model1-C-coefficients}.

\item \textbf{Model~2} contains the following input parameters at the GUT scale:
\begin{align}
y^d_{1},\ y^d_{2},\ y^d_{3},\ y^u_{1},\ y^u_{2},\ y^u_{3},\ \theta^{uL}_{12},\ \theta^{uL}_{23},\ \theta^{uL}_{13},\ \delta,\ \phi_1,\ \phi_2;\quad m_0,\ M_{1/2},\ A_{0},\ \tan\beta,\label{eq:model2-input-parameters}
\end{align}
where $y^{u,d}$, $\theta^{uL}$, $\delta$ and $\phi$ are used in constructions of Eqs.~\eqref{eq:model2-implementation-yd}--\eqref{eq:model2-implementation-yu-2}.
\end{itemize}
Both models consist of $16$ input parameters, $12$ of which construct the Yukawa sector, and $4$ parameters are intrinsic to the cMSSM: $m_0$, $M_{1/2}$, $A_{0}$ specify the mSUGRA boundary conditions at the GUT scale, and $\tan\beta$ is the ratio of the MSSM Higgs VEVs.

Note that the two phases $\phi_1$ and $\phi_2$ in both models do not change the SM prediction of fermion masses and mixing angles, since they get removed in the construction of the CKM matrix, while they do have an impact on the quasi-Yukawa sector and hence proton decay predictions. These phases are typically referred to as ``GUT phases'' in the literature \cite{Ellis:1979hy,Ellis:2019fwf}. 
\par
All initial parameters are real. Taking into account that singular values in a Takagi decomposition are non-negative, and all quark masses are non-vanishing, the parameters in the Yukawa sector are allowed to be in the following ranges (for both models):
\begin{align}
y^{d}_{12},y^{d}_{21},y^{d}_{22},\ y^{d}_{1},y^{d}_{2},y^{d}_{3},\ y^{u}_{1},y^{u}_{2},y^{u}_{3}&> 0,\\
\theta^{u}_{12},\theta^{u}_{23},\theta^{u}_{13}&\in[0,\pi/2],\\
\delta,\phi_1,\phi_2&\in [0,2\pi),\\
\varphi&\in [-\pi,\pi).
\end{align}
The cMSSM parameters can in principle also be varied in a fit. In this paper, however, our goal is an analysis which disentangles the flavor effects from the SUSY spectrum effects on proton decay. For this reason, we fix the cMSSM parameters to common example values in both models, where the SUSY threshold corrections allow for a good fit to (charged) fermion masses and the Higgs mass. We thus consider only a particular cMSSM benchmark point in our analyses.  
\par
To simplify the fitting procedure in the two models, we fix the gauge couplings at the GUT scale $\MGUT=2\cdot 10^{16}\GeV$: $g_3=0.698$, $g_2=0.697$ and $g_1=0.704$, which is consistent with the experimental values at low energies (assuming MSSM running above the SUSY scale of around $3\,\mathrm{TeV}$). Furthermore, since the effective triplet mass scale $\MEFF$ only affects the nucleon decay widths by an overall scaling, i.e.\ $\Gamma\propto1/(\MEFF)^2$, we can choose a fixed value $\MEFF=10^{19}\,\mathrm{GeV}$ in the following analyses, and can easily relate the results to any other value. Finally, in order for the RGE of the ratio $y_\mu/y_s$ of the 2nd family to start at $6$ at the GUT scale (based on our two models) and run to the experimental value at the scale $\MZ$, we choose the sign of the $\mu$ coupling (connecting $H_u$ and $H_d$) in the MSSM to be $\sign\mu=+1$, such that the dominant gluino loop diagram in SUSY thresholds drives $y_s$ in the correct direction \cite{Antusch:2009gu,Antusch:2008tf}.

\subsection{Observables}
\label{sec:observables}
Both models are fit to the same $14$ SM observables at low energies:
\begin{align}
y_u,\ y_c,\ y_t,\ y_d,\ y_s,\ y_b,\ \theta^\text{CKM}_{12},\ \theta^\text{CKM}_{23},\ \theta^\text{CKM}_{13},\ \delta^\text{CKM},\ y_e,\ y_\mu,\ y_\tau,\  m_h,\label{eq:model-observables}
\end{align}
corresponding to the masses of fermions in the up, down and charged lepton sector, the CKM parameters of the quark sector, and the SM Higgs mass $m_h$. 

Although there are $16$ input parameters in both models, $2$ GUT phases $\phi_1$ and $\phi_2$ have no influence on the listed observables, but only on nucleon decay discussed later, so the fit has effectively only $14$ relevant parameters. Since that number equals the number of listed observables, a reasonable expectation is that a good fit for both models can be obtained, with no flat directions in the $\chi^2$.

The values of the SM Yukawa observables at $\MZ$ and the error bars are taken from \cite{Antusch:2013jca}, where for the Yukawa couplings of the charged leptons we take a standard deviation of $1\,\%$ of the central value, which roughly corresponds to the precision of the running. The experimental value of $m_h$ is given in PDG (2018)~\cite{Tanabashi:2018oca}, where we take a standard deviation of $3\GeV$, namely $125.6\pm 3\GeV$, due to the theoretical uncertainties present in the calculation (for which we use FeynHiggs, see Section~\ref{sec:procedure}).

Other observables of interest predicted by the model are
the nucleon decay rates. We use our extended \texttt{SusyTC} code \texttt{SusyTCProton} to compute $13$ different decay channels, $8$ of which are proton decay channels and $5$ are neutron decay channels (we sum over the neutrino flavor states, since the decay experiments do not distinguish them); they are listed in Table~\ref{tab:decay-lifetimes} along with their experimental bounds (we took the most strict bound we found across multiple experiments). We shall not consider these observables in a $\chi^2$ fit,
since the decay rates themselves are only predicted up to an overall factor, unless the effective triplet mass parameter $\MEFF$ is also known. Taking a high enough $\MEFF$ can always avoid all proton decay bounds for $D=5$ contributions (we remind the reader that this parameter in our setup is not directly connected to any physical scale of triplets in the theory, so it can be taken above the Planck scale). Despite the lack of predictivity in nucleon rates themselves, our setup does allow for unambiguous predictions of branching ratios and, more generally, ratios of decay rates in different channels, e.g.~a ratio between a proton and neutron decay channel. 

\begin{table}[htb]
\begin{center}
\caption{ The $90\,\%$ confidence level exclusion bounds on proton and neutron decay from the Super-Kamiokande experiment for various ($B-L$)-preserving decay channels with a meson and a lepton in the final state. For each decay channel the corresponding lifetime bound $\tau/\mathcal{B}$ and the translated partial decay width $\Gamma_{\text{partial}}$ are listed, where $\mathcal{B}$ is the branching ratio for the decay channel. The data is taken from Figure~$5$-$3$ in~\cite{Brock:2012ogj} and further updated with other sources \cite{Takenaka:2020vqy,Bajc:2016qcc,Mine:2016mxy,Miura:2016krn,Abe:2014mwa,Regis:2012sn}.
\label{tab:decay-lifetimes}
}
\renewcommand{\arraystretch}{1.3}
\begin{tabular}{lrr}
\toprule
decay channel & $\tau/\mathcal{B}$ $[\text{year}]$ & $\Gamma_{\text{partial}}$ $[\mathrm{GeV}]$ \\
\midrule
Proton \\
\cline{1-1}
$p\rightarrow \pi^0\,e^+$ & $>2.4\cdot10^{34}$ & $<8.7\cdot10^{-67}$ \\
$p\rightarrow \pi^0\,\mu^+$ & $>1.6\cdot10^{34}$ & $<1.3\cdot10^{-66}$ \\
$p\rightarrow \eta^0\,e^+$ & $>4.1\cdot10^{33}$ & $<5.1\cdot10^{-66}$ \\
$p\rightarrow \eta^0\,\mu^+$ & $>1.2\cdot10^{33}$ & $<1.7\cdot10^{-65}$ \\
$p\rightarrow K^0\,e^+$ & $>1.1\cdot10^{33}$ & $<1.9\cdot10^{-65}$ \\
$p\rightarrow K^0\,\mu^+$ & $>1.6\cdot10^{34}$ & $<1.3\cdot10^{-66}$ \\
$p\rightarrow \pi^+\,\bar{\nu}$ & $>2.8\cdot10^{32}$ & $<7.4\cdot10^{-65}$ \\
$p\rightarrow K^+\,\bar{\nu}$ & $>6.6\cdot10^{33}$ & $<3.2\cdot10^{-66}$ \\
\midrule
Neutron \\
\cline{1-1}
$n\rightarrow \pi^-\,e^+$ & $>2.1\cdot10^{33}$ & $<1.0\cdot10^{-65}$ \\
$n\rightarrow \pi^-\,\mu^+$ & $>9.9\cdot10^{32}$ & $<2.1\cdot10^{-65}$ \\
$n\rightarrow \pi^0\,\bar{\nu}$ & $>9.9\cdot10^{32}$ & $<2.1\cdot10^{-65}$ \\
$n\rightarrow \eta^0\,\bar{\nu}$ & $>5.6\cdot10^{32}$ & $<3.7\cdot10^{-65}$ \\
$n\rightarrow K^0\,\bar{\nu}$ & $>1.2\cdot10^{32}$ & $<1.7\cdot10^{-64}$ \\
\bottomrule
\end{tabular}
\renewcommand{\arraystretch}{1}
\end{center}
\end{table}

\subsection{Calculational procedure\label{sec:procedure}}
For given values of the parameters in model~1 and 2, listed in Eq.~\eqref{eq:model1-input-parameters} and \eqref{eq:model2-input-parameters} respectively, the Yukawa matrices and the soft terms are implemented at the GUT scale $\MGUT=2\cdot10^{16}\GeV$ according to Section~\ref{sec:implementation}. 
Moreover, the dimension~$5$ operators which are relevant for nucleon decay are calculated at $\MGUT$ as well, using the formulas from Eq.~\eqref{eq:C5-in-model-begin} and \eqref{eq:C5-in-model-end}.
\par
The RG running in the MSSM is performed by means of the Mathematica package \texttt{SusyTCProton} (cf.~Appendix~\ref{sec:user_guide_SusyTCProton}), which is an extension of the Mathematica package \texttt{SusyTC}~\cite{Antusch:2015nwi} used within \texttt{REAP}~\cite{Antusch:2005gp}. The running of the Yukawa matrices, gauge couplings and soft terms is computed at $2$-loop, whereas the running of the dimension~$5$ operators is computed at $1$-loop.
\par
The SUSY scale $\MSUSY$ is determined dynamically in \texttt{SusyTCProton} by the geometric mean of the two up-type squark masses which have the largest mixing with the stop flavour eigenstates $\tilde{t}_1$ and $\tilde{t}_2$, namely $\MSUSY=\sqrt{m_{\tilde{t}_1}m_{\tilde{t}_2}}$. The values of the Yukawa matrices, gauge couplings and soft terms at $\MSUSY$ are used to calculate the partial decay widths of the proton and neutron listed in Table~\ref{tab:decay-lifetimes} by means of the Mathematica package \texttt{ProtonDecay} (cf.\ Appendix~\ref{sec:user_guide_ProtonDecay}). In addition, the electroweak (EW) Higgs mass is calculated by using \texttt{FeynHiggs~2.16.0}~\cite{Bahl:2018qog,Bahl:2017aev,Bahl:2016brp,Hahn:2013ria,Frank:2006yh,Degrassi:2002fi,Heinemeyer:1998np,Heinemeyer:1998yj}.
\par
In \texttt{SusyTCProton}, the MSSM is matched to the SM by taking the SUSY threshold corrections into account. Furthermore, the running from the SUSY scale to the $Z$-boson mass scale $\MZ=91.2\GeV$ is calculated at $2$-loop. At $\MZ$, the observables in the quark and charged lepton sector, namely the Yukawa couplings and the CKM parameters, are computed.
\par
For the statistical model analyses we use the following procedures:
\begin{itemize}
\item As a measure of goodness of fit we define the $\chi^2$ function in the usual way: 
\begin{align}
\chi^2(\vec{x})=\sum_{i}\frac{\big(f_i(\vec{x})-y_i\big)^{2}}{\sigma_{i}^2}\equiv \sum_{i} \chi^{2}_i,\label{eq:chi2-definition}
\end{align}
where the vector $\vec{x}$ consists of the input parameters of the models from either Eq.~\eqref{eq:model1-input-parameters} or \eqref{eq:model2-input-parameters}, and the index $i$ goes over all observables. The $y_i$ denote the central values from the (experimental) data, $\sigma_i$ are their corresponding standard deviation errors, while $f_i(\vec{x})$ are the predictions for the $i$-th observable given the parameter point $\vec{x}$. Some observables may be equipped with asymmetric errors $\sigma_{i+}$ and $\sigma_{i-}$ when $f_{i}(\vec{x})>y_i$ or $f_{i}(\vec{x})<y_i$, respectively. Each $\chi^{2}_i$ term in the $\chi^2$ function is referred to as a pull associated to a particular observable; it will also be convenient to refer to a particular pull by replacing the index $i$ with the label of the $i$-th observable. 

\item The best-fit point for a given model setup is calculated by using a differential evolution algorithm for global minimization using custom Mathematica code.

\item To calculate the posterior density of the parameters and observables for a certain model setup, we perform an MCMC analysis by using an adaptive Metropolis-Hastings algorithm~\cite{Roberts:2009}. All prior probability distributions for the parameters are chosen to be flat. Furthermore, we calculate $12$ independent chains with $5\cdot10^4$ valid points (beyond the ``burn-in phase''), which gives $6\cdot10^5$ points in total for each dataset. All chains are started at the best-fit point of the corresponding model setup.
\end{itemize}

\section{Results for nucleon decay from \texttt{SusyTCProton}\label{sec:results}}
In this section, we present the results of our numerical analyses for the two models we considered. Our main interest are the predicted nucleon decay rates for the points with a good fit to the low energy data.

Expressions for nucleon decay rates mediated by $D=5$ operators contain many contributions, all of which depend on both the masses of the SUSY sparticles that are used to dress the $D=5$ operators, as well as flavor effects both from the SM Yukawa sector and the sfermion sector. More details on the expressions used are given in Appendix~\ref{sec:documentation} and references specified there. 

The \texttt{SusyTCProton.m} and \texttt{ProtonDecay.m} packages developed for such a nucleon decay calculation, see Appendices~\ref{sec:user_guide_SusyTCProton} and \ref{sec:user_guide_ProtonDecay}, can handle all these effects. Our main interest in this paper are SUSY GUTs, and as such we are interested in the effect the Yukawa textures are having on proton decay. For this reason, we would like to isolate this effect from other contributions: taking cMSSM boundary conditions already insures no extra flavor violation in the SUSY sector, while the effects of variable sparticle masses can be removed by fixing the soft parameters.

We therefore choose in our analysis to fix the cMSSM parameters and consider only a benchmark point. We take the following values:
\begin{align}
\begin{split}
\tan\beta&=\phantom{+}28.4,\\
m_0&=\phantom{+0}4000\GeV,\\
M_{1/2}&=\phantom{+0}1700\GeV,\\
A_0&=-10000\GeV.\\
\end{split} \label{eq:cMSSM-benchmark-values}
\end{align}
We use these values in both models, and they allow us to obtain good fits for both\footnote{Both models have by design the same CG coefficient between the down and charged lepton sectors in the $3$-$3$ and $2$-$2$ entries at the GUT scale, and thus require essentially the same SUSY threshold corrections. A good fit to the fermion data is thus possible with the same values of soft parameters.}, thus being able to compare them on an equal footing and isolate the flavor effects. We have checked that EW vacuum stablility for this parameter point by using \texttt{Vevacious~1.2.03}~\cite{Camargo-Molina:2013qva}, \texttt{SPheno~4.0.3}~\cite{Porod:2003um,Porod:2011nf}, and the predefined model from \texttt{SARAH 4.14.1}~\cite{Staub:2008uz,Staub:2013tta} with possible charge breaking via stau VEVs. The output shows that the produced EW vacuum is stable.

Fixing the cMSSM parmeters leaves $12$ free parameters to vary for a fit, with 10 having an effect on the fit, while varying the GUT phases $\phi_1$ and $\phi_2$ will only effect the nucleon decay rates.

\subsection{Best-fit points and nucleon decay \label{sec:results-best-fit}}
Minimizing the $\chi^2$ with respect to the flavor parameters, we obtain a best-fit point for each of the models. The parameter values for the best-fit points are shown in Table~\ref{tab:results-bestfit-parameters}, while the $\chi^2$ results and various dominant pulls are shown in Table~\ref{tab:results-bestfit-pulls}. Most of the tension in the fit is coming from the pull in the $s$-quark mass.

\newcommand{\TEN}[1]{$\times 10^{-#1}$}

\begin{table}[ht]
\begin{center}
\caption{The flavor input parameters for the best-fit point of models 1 and 2. The GUT phases $\phi_1$ and $\phi_2$ have no influence on the predictions. \label{tab:results-bestfit-parameters}}
\begin{tabular}{llllllllll}
\toprule
\multicolumn{10}{l}{Model 1}\\
\midrule
$y^{u}_{1}$ & $y^{u}_{2}$ & $y^{u}_{3}$ & $\theta^{uL}_{12}$ & $\theta^{uL}_{23}$ & $\varphi$ & $y^{d}_{12}$ & $y^{d}_{21}$ & $y^{d}_{22}$ & $y^{d}_{3}$\\[6pt]
2.84 & 1.39 & 4.90 & 8.66 & 3.53 & $-4.49$ & 4.84 & 5.59 & 2.11 & 1.54 \\
\TEN{6} & \TEN{3} & \TEN{1} & \TEN{2} & \TEN{2} & \TEN{3} & \TEN{4}& \TEN{4}& \TEN{3}& \TEN{1}\\[6pt]
\midrule
\multicolumn{10}{l}{Model 2}\\
\midrule
$y^{u}_{1}$ & $y^{u}_{2}$ & $y^{u}_{3}$ & $\theta^{uL}_{12}$ & $\theta^{uL}_{23}$ & $\theta^{uL}_{13}$ & $\delta$ & $y^{d}_{1}$ & $y^{d}_{2}$ & $y^{d}_{3}$ \\[6pt]
2.85 & 1.39 & 4.90 & 2.27 & 3.45 & 7.40 & 5.88 & 1.24 & 2.19 & 1.54\\
\TEN{6} & \TEN{3} & \TEN{1} & \TEN{1} & \TEN{2} & \TEN{3}& & \TEN{4}& \TEN{3}& \TEN{1}\\
\bottomrule
\end{tabular}
\end{center}
\end{table}

\begin{table}[ht]
\begin{center}
\caption{The total $\chi^{2}$ and the dominant pulls $\chi^{2}_{i}$, as defined in Eq.~\eqref{eq:chi2-definition}, for the best-fit points of models 1 and 2 of Table~\ref{tab:results-bestfit-parameters}. The subdominant pulls $\chi^{2}_{i}<0.01$ are not shown. \label{tab:results-bestfit-pulls}}
\begin{tabular}{lr@{\hspace{1cm}}p{1cm}p{1cm}p{1cm}p{1cm}p{1cm}p{1cm}}
\toprule
model&$\chi^{2}$&$\chi^{2}_{y_{d}}$&$\chi^{2}_{y_{s}}$&$\chi^{2}_{y_{b}}$&$\chi^{2}_{\theta^{CKM}_{23}}$&$\chi^{2}_{\theta^{CKM}_{13}}$&$m_{h}$\\
\midrule
1&4.55 & 0.20 & 3.29 & 0.11 & 0.08 & 0.18 & 0.67\\[3pt]
2&5.92 & 0.01 & 4.79 & 0.11 & 0.14 & 0.18 & 0.67\\
\bottomrule
\end{tabular}
\end{center}
\end{table}

We can now use these best-fit points to obtain predictions of proton and neutron decay rates in various decay channels. We postpone a more comprehensive comparison of all the channels to Section~\ref{sec:results-MCMC}, and will only focus here on some selected and most interesting aspects of the results, namely the dependence of the rates and their ratios on the GUT phases $\phi_1$ and $\phi_2$, since these phases are not determined by the fit and we thus have an embedded ambiguity of their values in the models. We plot the results in this section as contour plots in the $\phi_{1}$-$\phi_{2}$ plane, and take a fixed value for the effective triplet mass $\MEFF=10^{19}\GeV$ where needed to obtain specific numerical results. Each of these plots is equipped with it's own scale legend to the right. Note that despite the same color range for contour regions, the values represented by these colors change between different plots. Also, some of the plots provide information about the location and value of the minimum and maximum, which are added to the plots in red color. This information of the minimum and maximum is important, since it indicates that the predictions robustly lie in a compact region, and the decay rates cannot be made to vanish by 
a suitable choice of $\phi_{1}$ and $\phi_2$. 

To draw the plots, we computed the values on an equidistant 2D-grid with increments $\pi/30$ for both the $\phi_1$ and $\phi_2$ directions, while a separate minimization/maximization routine was run to obtain the extrema in the $\phi_{1}$-$\phi_{2}$ plane.

\begin{figure}[htb]
\includegraphics[width=16cm]{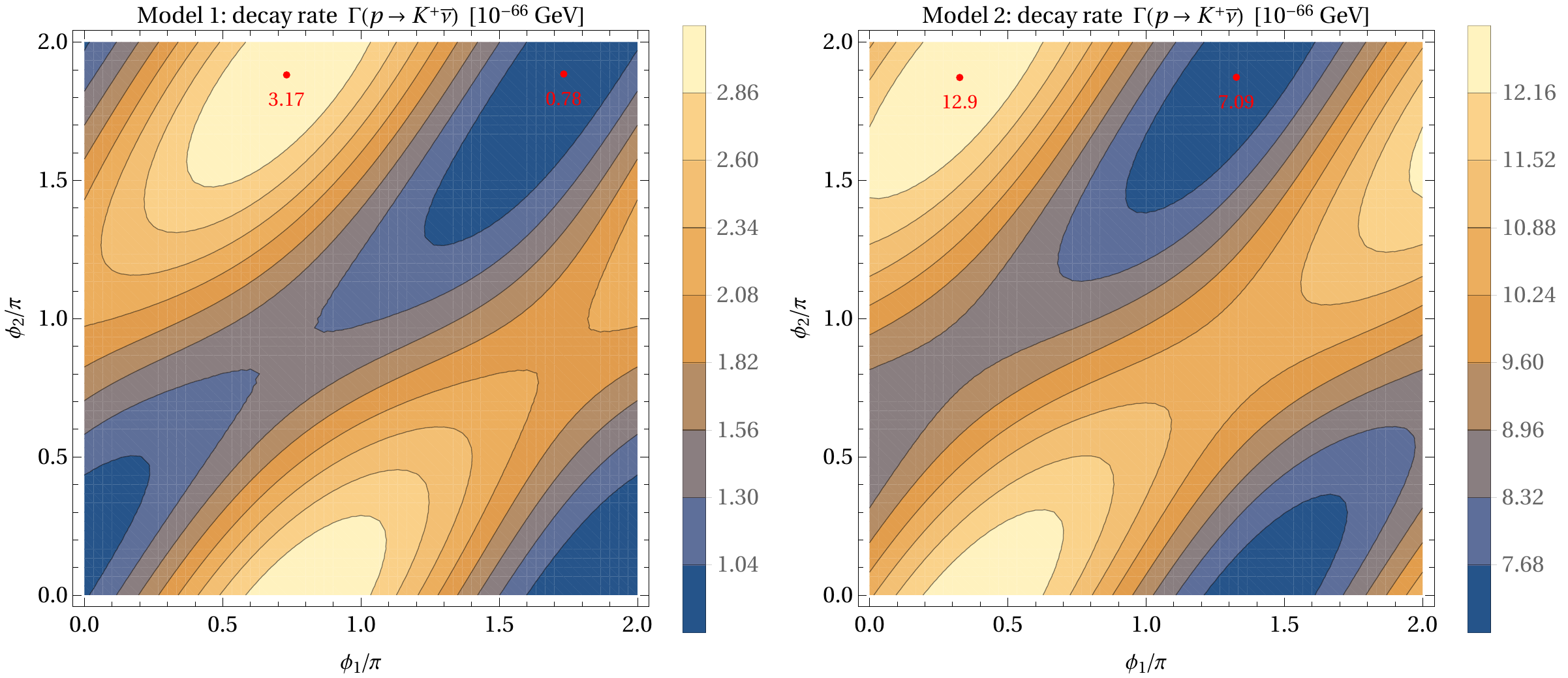}\\
\includegraphics[width=16cm]{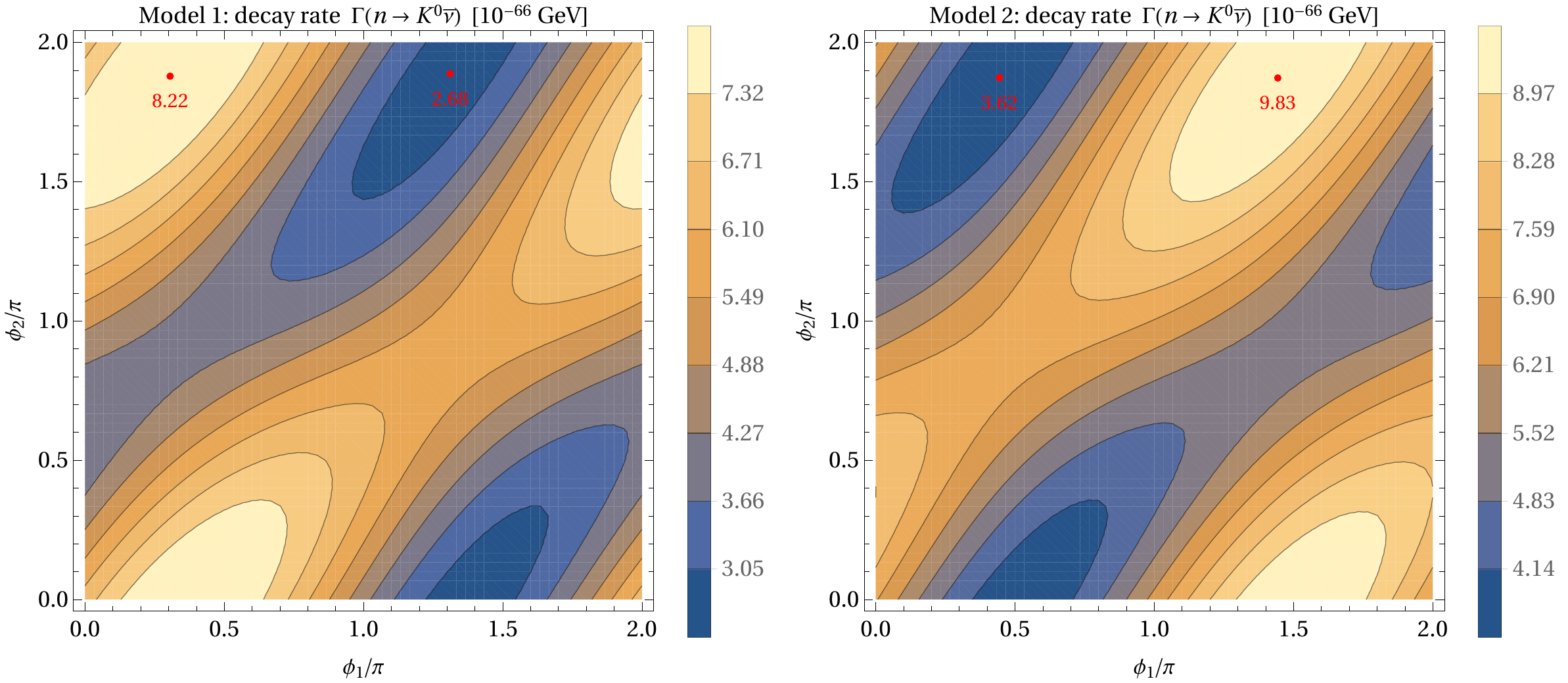}
\begin{center}
\caption{Contour plots for the GUT phase dependence of the dominant proton decay channel $p\to K^{+} \bar{\nu}$ (top) and neutron decay channel $n\to K^{0}\bar{\nu}$ (bottom) for the best-fit points of model 1 (left) and model 2 (right). We assumed $\MEFF=10^{19}\GeV$.\label{fig:phi-dependence-dominant-rates}}
\end{center}
\end{figure}

\begin{figure}[htb]
\includegraphics[width=16cm]{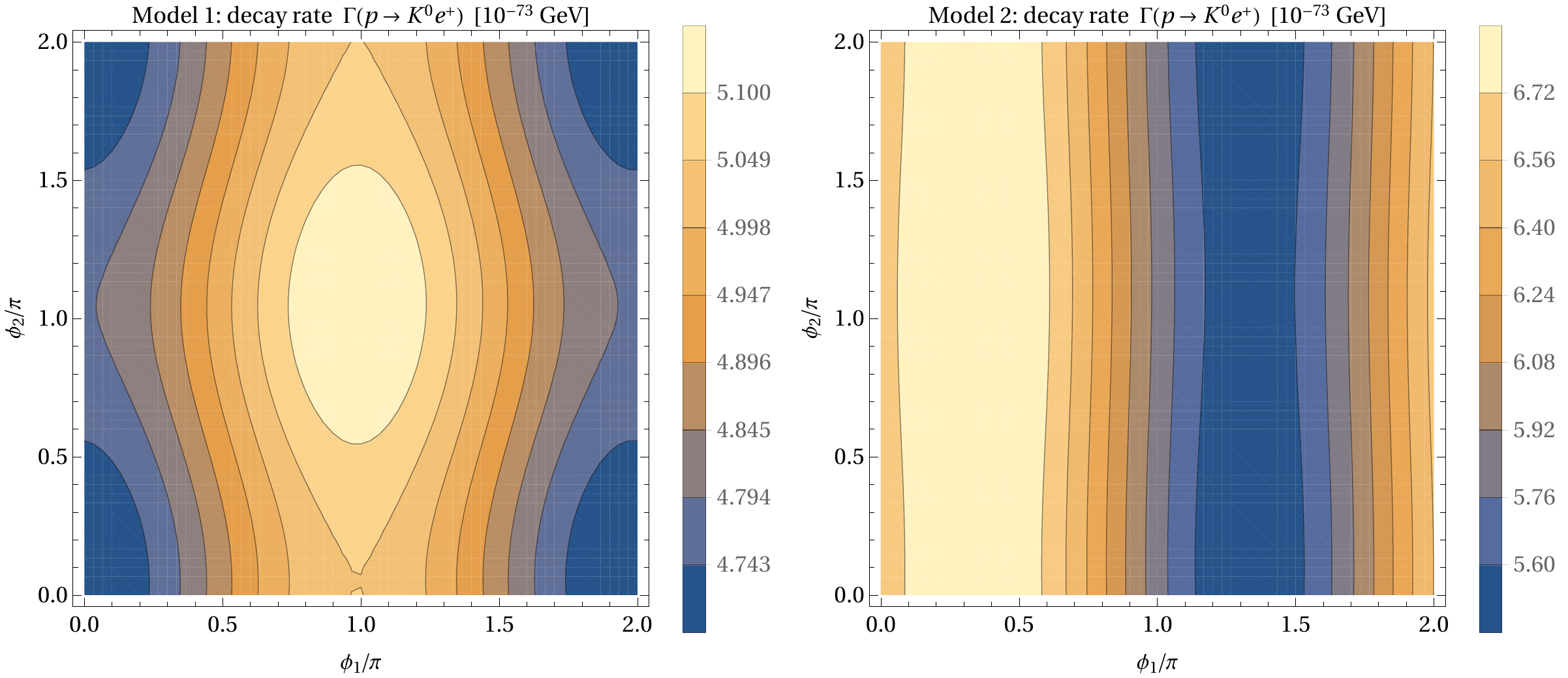}
\begin{center}
\caption{Contour plots for the GUT phase dependence of the $p\to K^{0} e^{+} $  proton decay channel for the best-fit points of model 1 (left) and model 2 (right). We assumed $\MEFF=10^{19}\GeV$.\label{fig:phi-dependence-selected-rates}}
\end{center}
\end{figure}

\begin{figure}[htb]
\includegraphics[width=16cm]{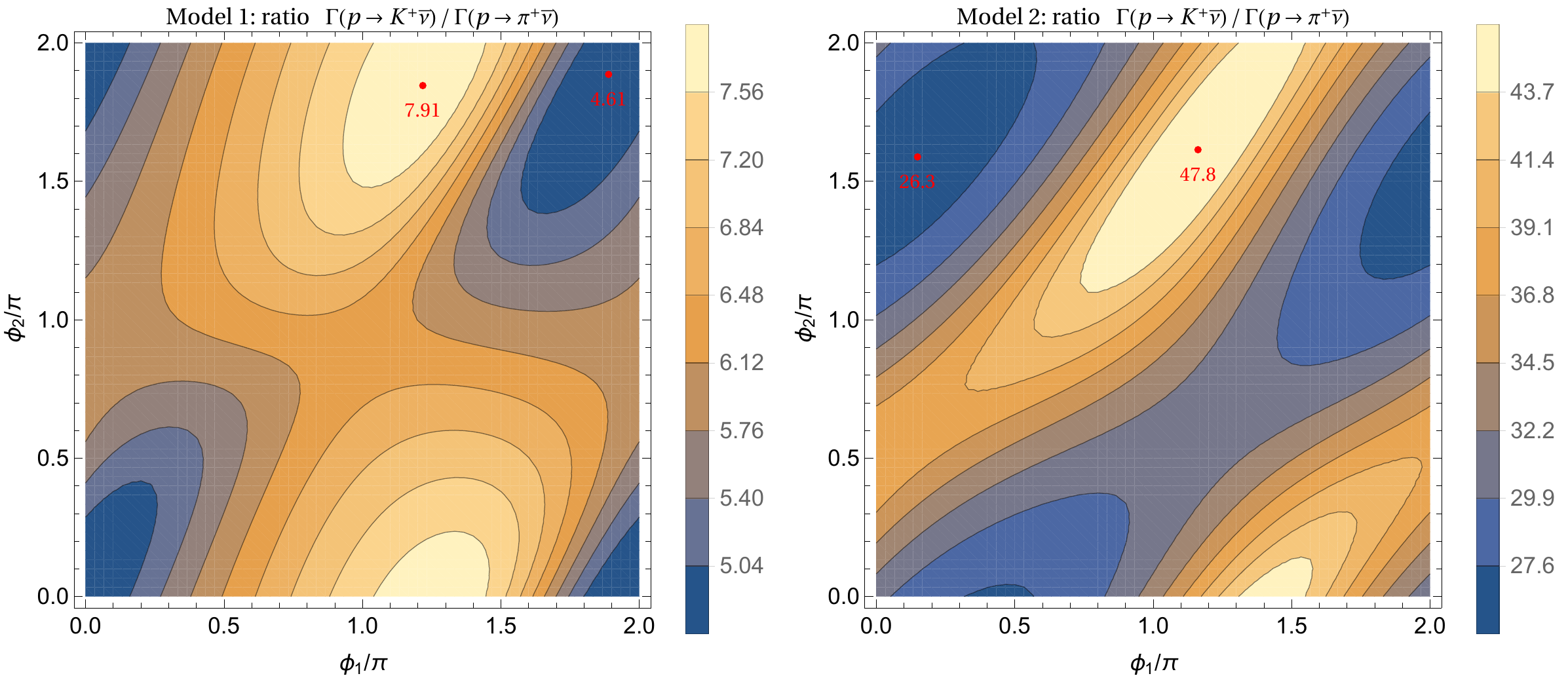}\\
\includegraphics[width=16cm]{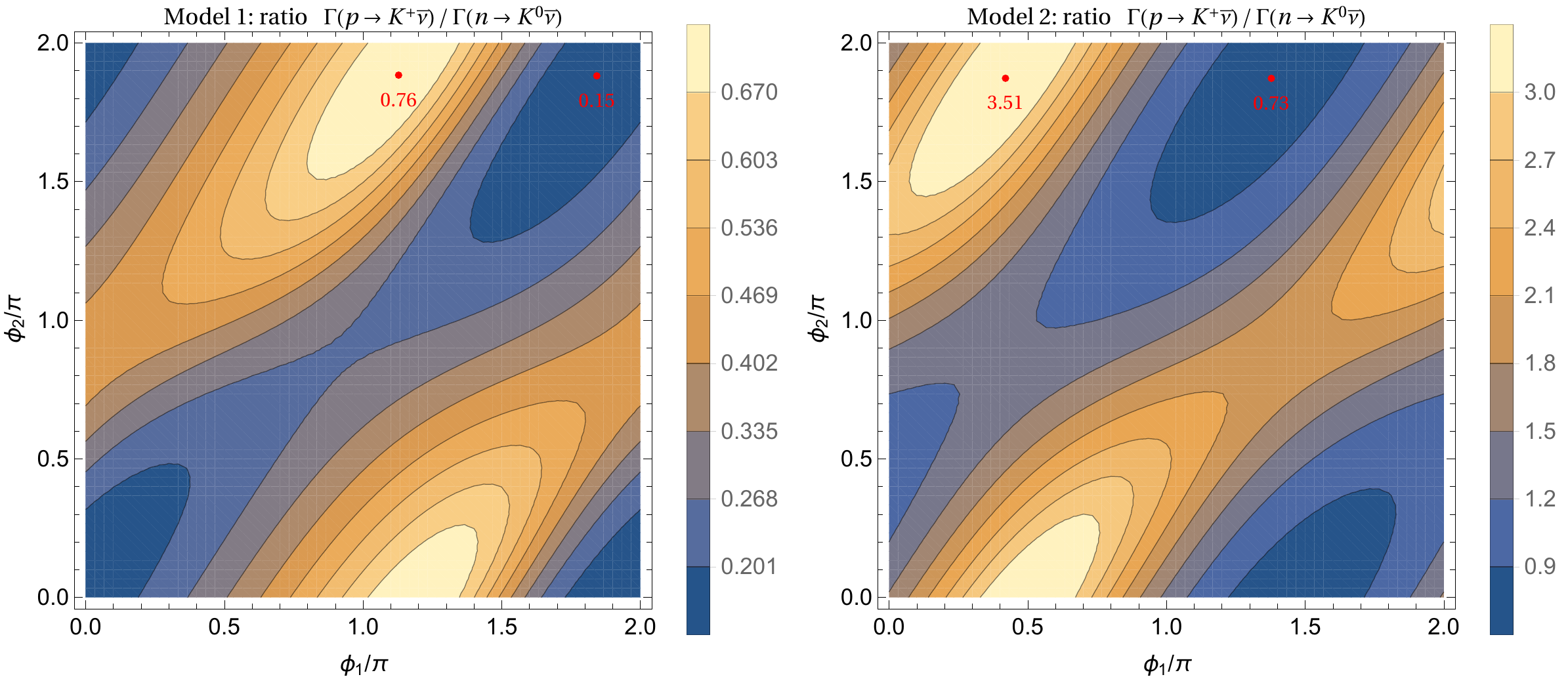}
\begin{center}
\caption{Contour plots for the GUT phase dependence of the decay rate ratios
\hbox{$\Gamma_{p\to K^{+}\bar{\nu}}/\Gamma_{p\to \pi^{+}\bar{\nu}}$} (top) and  \hbox{$\Gamma_{p\to K^{+}\bar{\nu}}/\Gamma_{n\to K^{0}\bar{\nu}}$} (bottom) for the best-fit points of model 1 (left) and model 2 (right). The results are independent of the effective triplet mass $\MEFF$.\label{fig:phi-dependence-ratios}}
\end{center}
\end{figure}

We make the following observations of the results:

\begin{itemize}
\item The dominant decay channels of both the proton and neutron are those with neutrinos in the final state; furthermore, the decay channel with a kaon always dominates over that of the pion. We plot the GUT phase dependence of the dominant proton and neutron decay rate for each model in Figure~\ref{fig:phi-dependence-dominant-rates}. 

We see that the rates for each process in a given model can differ by a factor of up to $\sim 3$, depending on the unknown GUT phases, but can never vanish completely. Observing the decay rate values in each panel, we see that the dominant neutron decay channel dominates over the proton decay channel in the case of model~1, while the opposite is true in model~2. Also, for model~1  the $\phi_{1}$-$\phi_{2}$ regions minimizing the proton decay rates roughly correspond to those minimizing the neutron decay rate; similarly, the proton and neutron decay rates are simultaneously maximal in a similar region. For model~2, on the other hand, minimizing the dominant proton decay roughly maximizes the neutron decay rate, and vice versa.
\item We selected to show another decay channel, $p\to K^{0}e^{+}$, in Figure~\ref{fig:phi-dependence-selected-rates}. The decay rates of this channel are highly suppressed compared to the dominant channels, and the plots show a much simpler GUT phase dependence. For model~2, where the charged lepton Yukawa
is diagonal, the rate effectively depends only on the phase $\phi_{1}$; in model~1, where the charged lepton mass matrix is not diagonal, the relative changes of the rates are smaller, but the $\phi_{2}$ dependence is not negligible. 
\item A result independent of the effective triplet mass can be obtained by considering ratios of decay rates. We choose to present two such ratios in the plots: the decay ratio $\Gamma_{p\to K^{+}\bar{\nu}}/\Gamma_{p\to \pi^{+}\bar{\nu}}$ represents the ratio between the two most dominant proton decay channels, while the ratio $\Gamma_{p\to K^{+}\bar{\nu}}/\Gamma_{n\to K^{0}\bar{\nu}}$ is between the dominant proton and neutron channels. The GUT phase dependence of the ratios is shown in Figure~\ref{fig:phi-dependence-ratios}.

We observe that the plots for ratios have a similar feature to those for the dominant rates: the ratios are simultaneously minimized or maximized for model~1, but minimizing one ratio roughly maximizes the other in model~2. This implies that the process $p\to \pi^{+}\bar{\nu}$, which is not shown in Figure~\ref{fig:phi-dependence-dominant-rates}, has a similar minimizing/maximizing region to the dominant proton decay process; the switch of the regions happens for model~2 only when migrating to neutron decays.

Another important observation is that the predicted range for a given ratio does not overlap when comparing the two models. This implies that experimentally determining such a ratio of nucleon decay channels would enable us to discriminate between the two flavor GUT models, despite the unknown GUT phases and despite the two models having the same soft SUSY-breaking parameters. This difference in the predicted ratios is thus purely a flavor effect, as could be expected; we provide additional discussion of using ratios for discrimination in Section~\ref{sec:results-MCMC}.   
\end{itemize}

\subsection{MCMC results for nucleon decay \label{sec:results-MCMC}}
The nucleon decay results of the previous section involved varying only the two GUT phases $\phi_{1}$ and $\phi_{2}$, while we keep the other flavor parameters
confined to the best-fit point in each model. 

We now analyze the two models by varying all flavor parameters of Eq.~\eqref{eq:model1-input-parameters} and \eqref{eq:model2-input-parameters}, such that the fit is still close to the best-fit point; the cMSSM prameters are still fixed, however, to their benchmark values of Eq.~\eqref{eq:cMSSM-benchmark-values}. For this analysis we use the Markov Chain Monte Carlo (MCMC) method, which allows us to obtain estimates of posterior probability densities on the parameter space. This results in obtaining highest posterior density (HPD) intervals for the observables of interest. For more technical details on our MCMC procedure see Section~\ref{sec:procedure} (and references there-in).

The plots in this section show HPD intervals of nucleon decay rates or ratios predicted by the two models. The HPD intervals are colored red (green) for model 1 (2). We show intervals corresponding to the $1$-$\sigma$ and $2$-$\sigma$ HPD regions as darker and lighter colored, respectively, and overlayed one on top of the other.

\begin{figure}[htb]
\begin{center}
\includegraphics[width=16.5cm]{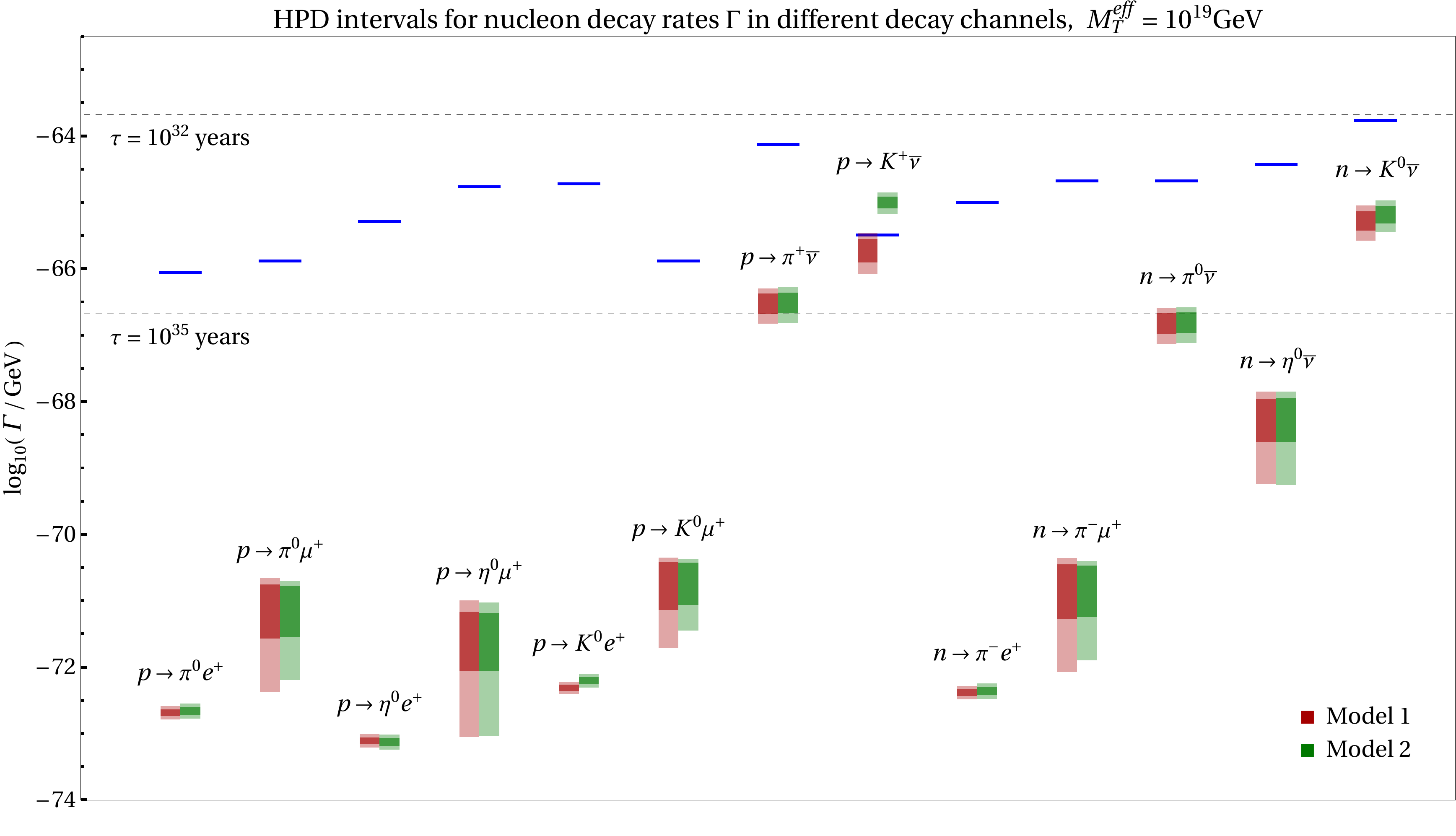}
\caption{The $1$-$\sigma$ (dark) and $2$-$\sigma$ (light) HPD regions predicted via MCMC for nucleon decay rates in both models using a benchmark cMSSM point and a fixed effective triplet mass $\MEFF=10^{19}\GeV.$ The blue line segments indicate current experimental bounds. \label{fig:sigma-ranges-rates}}
\end{center}
\end{figure}

\begin{figure}[htb]
\begin{center}
\includegraphics[width=16cm]{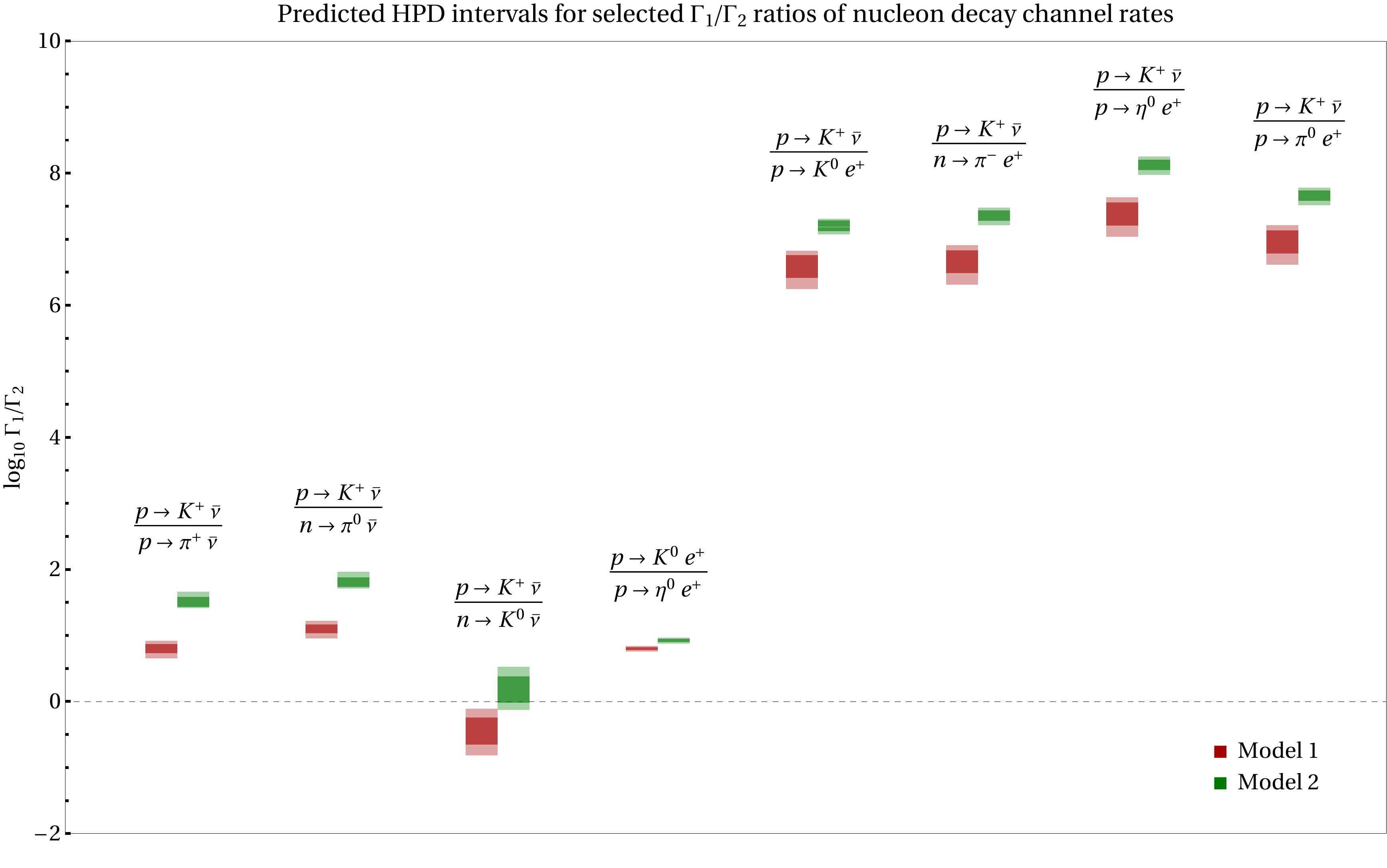}
\caption{The $1$-$\sigma$ (dark) and $2$-$\sigma$ (light) HPD regions predicted via MCMC for selected ratios of decay rates of two channels; we assume the same cMSSM benchmark point in both models. \label{fig:sigma-ranges-ratios}}
\end{center}
\end{figure}

We make the following observations of the results:
\begin{itemize}
\item Our \texttt{SusyTCProton} and \texttt{ProtonDecay} packages are able to compute $13$ different decay channels (specified in Table~\ref{tab:decay-lifetimes}). The MCMC results for the HPD intervals in each channel are shown in Figure~\ref{fig:sigma-ranges-rates}. Note that we assumed the effective triplet mass to be $\MEFF=10^{19}\GeV$; the decay rate results for a different $\MEFF$ can be easily obtained by noting that $\Gamma\propto (\MEFF)^{-2}$, implying that we linearly translate the intervals down by $2$ on the log scale for each order of magnitude that $\MEFF$ increases over the reference value. The experimental bounds for the various channels are shown as blue bars; choosing a different $\MEFF$ can control whether we violate any of the experimental bounds or not with the predictions. For the reference value of $\MEFF$, the experimental bound on $p\to K^{+}\bar{\nu}$ would exclude model~2 but not model~1.  

We see that the dominant channels are those with a neutrino in the final state, while those with a charged lepton have decay rates at least $4$ orders of magnitude below the dominant ones. The dominant proton decay channel is $p\to K^{+}\bar{\nu}$, while the dominant neutron decay channel is $n\to K^{0}\bar{\nu}$. Which of the two is bigger differs between the two models. This result is consistent with the dominant decay channels when only the GUT phases were varied in Section~\ref{sec:results-best-fit}.

Also, we notice that when a charged lepton is a decay product, the predictions for the rates are much sharper when the charged lepton is $e^{+}$ (the $2$-$\sigma$ ranges vary $\sim 20\,\%$ around the average), while the prediction for $\mu^{+}$ in the final state typically spans $1$ to $2$ orders of magnitude. Given the decay rates of these processes, however, they are at least $4$ orders of magnitude below the dominant channels and thus experimentally not within reach. 

Comparing the HPD intervals between the two models process by process, we notice that for a given $\MEFF$ most channels are predicted with a similar decay rate, except for $p\to K^{+}\bar{\nu}$; this process is thus most suitable for discrimination between the two models, assuming another channel has been measured as well. Luckily, the decay $p\to K^{+}\bar{\nu}$ also has either the biggest or 2nd biggest decay rate, so it would likely be among the first experimentally measured channels if one of the considered models is realized in nature.
\item The quantities independent of $\MEFF$ are the ratios of various decay rates. Note that a ratio for each parameter point in the MCMC dataset is computed separately, and only then is statistics performed on these values; this is an important point, since there could be additional correlations between various decay channels in a model not apparent when viewing only the HPD intervals of decay rates, i.e.~when considering the numerator and denominator in the ratio separately. 

In Figure~\ref{fig:sigma-ranges-ratios} we show selected ratios which best discriminate between the two models. They effectively have no overlap of predicted $2$-$\sigma$ HPD intervals when comparing the two models, thus measuring even one such ratio has sufficient discriminating power. All ratios between the $13$ channels not shown in the Figure all have some overlap in the $2$-$\sigma$ intervals. More broadly, given any specific model, predictions for a set of ratios effectively provides a fingerprint of that model, in principle allowing for it to be excluded if multiple channels of decays were measured experimentally. 

The most interesting ratios experimentally are those involving the dominant decay channels, which would be the first ones measured. Also, a very big or small prediction of the ratio (compared to a $1:1$ ratio) implies a very big discrepancy in the rates of the numerator and denominator, again making it unlikely that both processes could be measured in the near future. Given these considerations, the most experimentally relevant ratios are the first three, which compare the high rate processes with a neutrino in the final state. 

All but one of the ratios in Figure~\ref{fig:sigma-ranges-ratios} include the $p\to K^{+}\bar{\nu}$ decay as one of the two processes, confirming our comment for Figure~\ref{fig:sigma-ranges-rates} that this process is a good discriminator between the two models. In terms of the sharpness of the prediction of ratios, the value for the fourth ratio in the figure has the smallest allowed range: the $2$-$\sigma$ HPD intervals for \hbox{$\Gamma_{p\to K^{0}e^{+}}/\Gamma_{p\to\eta^{0}e^{+}}$} are $[5.7,6.9]$ and $[7.5,9.3]$ for models 1 and 2, respectively. The decay channels in that ratio have very small rates and are experimentally out of reach in the considered scenarios; assuming an experimental measurement of such a ratio (or even just one channel and having bounds on the other) provides, however, an excellent way to exclude the model under consideration.

A final note on ratios: some ratios are predicted exactly the same with a value $(1/\sqrt{2})^{\pm 1}$ for every point due to the form of the low-energy effective Lagrangian in chiral perturbation theory. In particular, these are the $3$ ratios where only a replacement of $\pi^{0}$ with $\pi^{\pm}$ takes place in the final state:
\begin{align}
\frac{\Gamma_{p\to \pi^{0} e^{+}}}{\Gamma_{n\to \pi^{-}e^{+}}},\quad
\frac{\Gamma_{p\to \pi^{0}\mu^{+}}}{\Gamma_{n\to\pi^{-}\mu^{+}}},\quad
\frac{\Gamma_{p\to\pi^{+}\bar{\nu}}}{\Gamma_{n\to \pi^{0}\bar{\nu}}}.
\end{align}
These ratios are fixed and have no dependence on any high-energy or SM parameters.
\end{itemize}

\newpage
\section{Conclusions \label{sec:conclusions}}

While the observation of nucleon decay would be a smoking gun of Grand Unified Theories (GUTs) in general, the ratios between the decay rates of the various channels carry rich information about the specific GUT model realization.
We investigated in this paper these nucleon decay fingerprints of two flavor SUSY GUT models based on the group $\mathrm{SU}(5)$. The two models differ in the texture of their Yukawa matrices, and the types of GUT operators generating the SM fermion masses, cf.~Section~\ref{sec:model_setup}. Both admit essentially equally good fits to the low energy Yukawa data of SM masses and mixing angles (with a $\chi^{2}\approx 5$, cf.~Section~\ref{sec:results-best-fit}), even given the same sparticle spectrum. Yet their nucleon decay computation shows that the partial decay rates of various channels are good observables to discriminate between them, despite the decay ambiguity introduced by $2$ unknown GUT phases. 

We investigated the dependence of the decay rate on the GUT phases at the best-fit point for the flavor parameters of the model in Section~\ref{sec:results-best-fit}.
The nucleon decay fingerprinting has been performed by computing the decay rates in $8$ different proton decay channels and $5$ different nucleon decay channel, cf.~Table~\ref{tab:decay-lifetimes}. We obtained their predicted ranges by varying the flavor parameters in a way still compatible with a low $\chi^2$ of the low-energy observables using the MCMC algorithm. The main results of the analysis are presented in Section~\ref{sec:results-MCMC}, see especially Figures~\ref{fig:sigma-ranges-rates} and \ref{fig:sigma-ranges-ratios}. The general features are those expected in a SUSY GUT: the $p\to K^{+}\bar{\nu}$ and $n\to K^{0}\bar{\nu}$ channels are the dominant modes of proton and neutron decay, respectively, and the rates with kaons in the final states are bigger than those with pions when the other decay product is the same. Other features found in the fingerprints are that the neutrino decay channels dominate over those with charged leptons in the final state and that the channels with $e^{+}$ in the final state are predicted much more sharply than those with $\mu^{+}$. 

The decay rates invariably depend on the scale of the heavy color triplets that mediate $D=5$ nucleon decay. We considered a doublet-triplet setup 
where the overall scale of the rates depends only on the triplet effective mass parameter $\MEFF$. Robust and $\MEFF$-independent predictions for both models are obtained by considering ratios of decay rates between different channels. The key connection of nucleon decay to flavor is that the couplings of the SM fermions to the $D=5$ nucleon decay mediating triplets are related to the Yukawa coupling of the theory, since they arise from the same GUT operators, which is a generic feature of any GUT realization. The results for the considered models show that the impact of their different flavor structure is already sufficient to discriminate between them. In particular, the dominant proton decay mode $p\to K^{+}\bar{\nu}$ turns out to differ relative to other dominant modes with neutrinos in the final states such as $p\to \pi^{+}\bar{\nu}$ and $n\to \pi^{0}\bar{\nu}$.

The computations of nucleon decay rates for the $13$ channels in this paper were performed by use of the Mathematica software modules \texttt{SusyTCProton} and \texttt{ProtonDecay}, which we developed for the purposes of this paper and are rolling them out for public use. The \texttt{SusyTCProton} is an extension of the \texttt{SusyTC} package within the \texttt{REAP} framework, and allows to compute the RG running of $D=5$ operators from the GUT scale to the $Z$ scale alongside the running of all other parameters in a softly broken MSSM, performing the MSSM to SM matching and dressing the $D=5$ operators at the SUSY scale. The \texttt{ProtonDecay} module admits as input either the \texttt{SusyTCProton} output or an SLHA file amended with $D=5$ operator values at the SUSY scale, and computes the resulting nucleon decay rates in the various channels. The module is able to include $D=6$ contributions to nucleon decay if the operator values are provided at some high scale where the SM is taken as a valid effective theory, so the package can be used for computing nucleon decay in non-SUSY GUTs as well. We list the documentation for the use of these software packages (a ``User's guide'') in Appendix~\ref{sec:user_guide}.

When used in tandem, the above software packages allow for the computation of the decay rates in any softly broken SUSY GUT if one is able to input the parameter values at a scale where the MSSM becomes valid (in our case this was immediately below the GUT scale). It computes the rates in all generality, including all the flavor effects from both the Yukawa and SUSY sectors, and considers all diagrams in these decays following \cite{Goto:1998qg}. This allows for the fingerprinting of nucleon decay within such a model, opening up a new avenue for the model to be tested in experiments. The fingerprint is most robust when considered at the level of ratios of decay rates for various channels. Experimentally, the measurement of decay events in just one channel either for the proton or neutron, alongside experimental bounds in the other channels, already allows for strategies for cross-checking with the fingerprint, since the model prediction might suggest that decay events in some other channels should have been observed already as well. A situation with decay events measured in multiple channels of course allows for even better tests of the model. These fingerprinting techniques can be readily used either to discriminate between otherwise indistinguishable models from  the point of view of current experiments, or to possibly rule them out.

This paper represents a thorough analysis of nucleon decay in two flavor GUT toy models, demonstrating the efficacy of using their nucleon decay fingerprints to distinguish them. Alongside this analysis, we also roll out new software tools which can be employed by the community for further investigation of nucleon decay in models of various types, thus adding to the nucleon decay fingerprint collection.

\section*{Acknowledgements}
This work has been supported by the Swiss National Science Foundation. The authors would like to thank Liliana Velasco-Sevilla for discussion.

\appendix
\appendixpage

\section{Details on doublet-triplet splitting\label{sec:DT-splitting}}

DT-splitting \cite{Dimopoulos:1981zb} needs to be addressed in any GUT theory. In the SUSY setting, the main issue is the following dichotomy:
\begin{itemize}
\item[{[D]:}] One mass-eigenpair of a doublet $D\sim (\mathbf{1},\mathbf{2},+1/2)$ and anti-doublet $\overline{D}\sim (\mathbf{1},\mathbf{2},-1/2)$ is identified with $H_u\oplus H_d$ of the MSSM, and their mass should be negligible compared to the GUT scale in order for EW symmetry breaking to occur. MSSM extensions with multiple doublets are of course possible, but we consider the MSSM as the effective theory below $\MGUT$ in this paper.
\item[{[T]:}] The triplet states $T\sim(\mathbf{3},\mathbf{1},-1/3)$ and anti-triplet states $\overline{T}\sim(\mathbf{\bar{3}},\mathbf{1},+1/3)$ can mediate dimension $5$ proton decay, so they should be heavy (at the GUT scale).
\end{itemize}

The two conditions are difficult to fulfill simultaneously due to irreducible representations of he unified group always containing a triplet alongside a doublet. Even in $\SU(5)$, which is the smallest possible unified group containing the SM and a subgroup of all other (true) unified group candidates, both doublets and triples are present in the representations $\mathbf{5}$ and $\mathbf{45}$ (and their conjugates). The only (low-dimensional) exception is the representation $\mathbf{50}$ of $\SU(5)$, which contains a triplet but no doublet; this feature is used in the missing partner mechanisms (see e.g.~\cite{Masiero:1982fe}) to achieve DT-splitting.

Since $D=5$ proton decay is mediated by heavy triplets, it unavoidably depends on the details of DT splitting, which in turn depends on the breaking sector of the theories. We wish to be as model independent in the breaking sector as possible for our nucleon decay analysis in this paper; for this reason we assumed the simplest possible scenario where only one triplet couples to the fermionic representations. As discussed in Section~\ref{sec:general-setup}, this implies that only the $1$-$1$ element of the inverse mass matrix $\mathbf{M}_T^{-1}$ is important for proton decay. 

This choice represents a minimal parametrization 
of DT-splitting with only one parameter: the effective triplet mass $\MEFF$, cf.~Eq.~\eqref{eq:effective-triplet-definition}. Due to Eq.~\eqref{eq:C5-in-model-begin}-\eqref{eq:C5-in-model-end}, the amplitude for all $D=5$ nucleon decay channels is proportional to $(\MEFF)^{-1}$, and the decay rates are proportional to $(\MEFF)^{-2}$. Raising the effective triplet mass can suppress the overall proton decay while keeping the decay rate ratios fixed (ignoring $D=6$ contributions).

In this appendix section we present two concrete realizations of such a scenario; despite multiple possibilities for the underlying dynamics, only the effective mass $\MEFF$ is the relevant parameter. Note that while this parameter has mass dimension $1$, it does not (necessarily) represent a physical mass of a particle or a characteristic energy scale at which a process takes place, and as such can take trans-Planckian values. Indeed, for our numerical analysis in Section~\ref{sec:results} we took for the effective triplet mass the benchmark value $\MEFF=10^{19}\GeV$, so the proposed mechanisms should be able to accommodate such a high parameter value.

We present two possibilities how to achieve a setup admitting high $\MEFF$: a modification of the ordinary missing partner mechanism is discussed in Appendix~\ref{sec:tweak}, while the double missing partner mechanism is described in Appendix~\ref{sec:double-missing-partner}.

\subsection{The modified missing partner mechanism\label{sec:tweak}}

\subsubsection{The missing partner mechanism ---  the usual setup}
In the usual setup, the missing partner mechanism in $\SU(5)$ \cite{Masiero:1982fe,Grinstein:1982um} requires that the color triplets $T$ (or anti-triplets $\overline{T}$) come from the following $\SU(5)$ representations:
\begin{align}
\mathbf{5}_H\oplus\mathbf{\bar{5}}_H\oplus\mathbf{50}_H\oplus\mathbf{\overline{50}}_H. \label{eq:DT-sector}
\end{align}
The Higgs doublets $(\mathbf{1},\mathbf{2},\pm 1/2)$ on the other hand are present only in the $\mathbf{5}_H\oplus\mathbf{\bar{5}}_H$ part; as mentioned earlier in Appendix~\ref{sec:DT-splitting}, the representations $\mathbf{50}$ ($\overline{\mathbf{50}}$) of $\SU(5)$ remarkably contains only a triplets (anti-triplet), but not an (anti-)doublet.

The doublet and anti-doublet directly correspond to the MSSM doublets $H_u$ and $H_d$, respectively:
\begin{align}
H_u&\subseteq \mathbf{5}_H,& H_d&\subseteq \mathbf{\bar{5}}_H.\label{eq:doublets}
\end{align}
The (anti-)triplets in Eq.~\eqref{eq:DT-sector} on the other hand are labeled by 
\begin{align}
T_1&\subseteq \mathbf{5}_H,&\overline{T}_1 &\subseteq \mathbf{\bar{5}}_H,\label{eq:triplets1}\\
T_2&\subseteq \mathbf{50}_H,&\overline{T}_2 &\subseteq \mathbf{\overline{50}}_H.\label{eq:triplets2}
\end{align}
We assume then an $\SU(5)$-level superpotential of the form
\begin{align}
W_{DT}&=\lambda\;\mathbf{5}_H\cdot\mathbf{\overline{50}}_H\cdot \mathbf{75}_H+\bar{\lambda}\;\mathbf{\overline{5}}_H\cdot\mathbf{50}_H\cdot \mathbf{75}_H
+m_{50}\;\mathbf{50}_H\cdot\mathbf{\overline{50}}_H,\label{eq:mp-W}
\end{align}
where the $G_{321}$ singlet component in $\mathbf{75}_H$ obtains a GUT scale VEV
$v=\langle\mathbf{75}\rangle$.\footnote{Alternatively, one could avoid the use of the representation $\mathbf{75}$ by utilizing a symmetric product of a pair of $\mathbf{24}$s instead.} Crucially, the explicit mass term $\mathbf{5}_H\cdot\mathbf{\overline{5}}_H$ is missing (this can be achieved for example by imposing discrete symmetries). The superpotential of Eq.~\eqref{eq:mp-W} yields the mass terms 
\begin{align}
\mathbf{W}\supset H_u\;\mathbf{M}_D\;H_d+
\begin{pmatrix}
T_{1}&T_{2}\\
\end{pmatrix}
\;\mathbf{M}_T\;
\begin{pmatrix}
\overline{T}_{1}\\ \overline{T}_{2}\\
\end{pmatrix}+\cdots,
\end{align}
where the doublet and triplet mass matrices $\mathbf{M}_D$ and $\mathbf{M}_T$ are
given by
\begin{align}
\mathbf{M}_{D}&=0,&
\mathbf{M}_{T}&=
\begin{pmatrix}
0&\lambda v\\
\bar{\lambda} v& m_{50}\\
\end{pmatrix}.\label{eq:missing-partner-MT}
\end{align}

In general we have $\lambda,\bar{\lambda},v,m\in\mathbb{C}$, but the phases of $\lambda v$, $\bar{\lambda} v$ and $m$ can be absorbed into the overall phase definition of the irreps $\mathbf{\overline{50}}_H$, $\mathbf{75}_H$ and $\mathbf{50}_H$ in Eq.~\eqref{eq:mp-W}, respectively. This implies that all entries in $\mathbf{M}_T$ can be taken to be real numbers. 

This situation in Eq.~\eqref{eq:missing-partner-MT} clearly achieves DT-splitting, since the doublet remains effectively massless, while all triplets are roughly at the GUT scale due to the presence of the VEV $v$. Furthermore, using Eq.~\eqref{eq:effective-triplet-definition}, such an $\mathbf{M}_T$ gives the effective triplet mass to be 
\begin{align}
\MEFF&=-\lambda\bar{\lambda}\,v^2/m_{50}.\label{eq:meff-missing-partner}
\end{align}

\subsubsection{The perturbativity problem and modifying the usual setup}

Clearly, $\MEFF$ from Eq.~\eqref{eq:meff-missing-partner} can be enhanced by taking a small value for the parameter $m_{50}$, i.e.~\hbox{$m_{50}\ll \MGUT$.} Although the scale of the triplet masses is not impacted, since those are controlled by the VEV $v$, the remainder of $\mathbf{50}\oplus\overline{\mathbf{50}}$ (consisting of particles of $1\times 1$ mass matrices, since they are not found in $\mathbf{5}_H\oplus\overline{\mathbf{5}}_H$) is then moved to the scale of $m_{50}$. 

This turns out to present a problem, since introducing a lot of matter below $\MGUT$ may increase the gauge coupling value to such a degree at $\MGUT$ that the Landau pole of the coupling is unacceptably close to the unification scale. Let's  investigate this situation: 
\begin{enumerate}
\item  The solution to the 1-loop RG running of the gauge coupling $g$ between two scales $\mu_r$ and $\mu'_r$ is described by 
\begin{align}
\alpha^{-1}(\mu_r')=\alpha^{-1}(\mu_r)+\tfrac{\beta}{2\pi}\log(\mu_r'/\mu_r),\label{eq:RGE-solution}
\end{align}
where $\alpha=g^2/(4\pi)$ and $\beta$ depends on the model. We remind the reader that in SUSY theories $\beta=3 D_{2}(V)-D_{2}(C)$, where $D_{2}$ is the Dynkin index, with $V$ and $C$ referring  to all vector and chiral supermultiplets in the theory with masses below a given renormalization scale, respectively. Adding a lot of matter (chiral supermultiplets) makes $\beta$ smaller (more negative), causing $g$ to blow up sooner when running to higher scales. 
\item Suppose we change the theory by moving some representations from the scale $\mu_0$ to $\mu_0/c$ ($c>1$), thus changing the $\beta$-function between $\mu_0/c$ and $\mu_0$ from $\beta$ to $\beta'$. The new coupling $\alpha'^{-1}$ and the old coupling $\alpha^{-1}$ should be the same at $\mu_0/c$, and using Eq.~\eqref{eq:RGE-solution} we can relate them at the scale $\mu_0$ via
\begin{align}
\alpha'^{-1}(\mu_0)&=\alpha^{-1}(\mu_0)+\tfrac{\beta'-\beta}{2\pi}\log c,\label{eq:g-shift}
\end{align}
which depends on the difference of the two beta functions $\Delta\beta=\beta'-\beta$. For simplicity, we consider the unified coupling and a setup where an entire $\mathbf{50}\oplus\overline{\mathbf{50}}$ is moved from $\mu_0=\MGUT$ to $\mu_0/c$: since it is a complete $\SU(5)$ representation, it does not change the unification scale $\mu_0$, but only the value of the unified coupling from $\alpha^{-1}(\mu_0)$ to $\alpha'^{-1}(\mu_0)$. 
\item
The scale of the Landau pole $\mu_L$ is defined implicitly by $\alpha^{-1}(\mu_L)=0$, i.e.~when $g$ has a pole. We define $C$ as the factor above the GUT scale where the Landau pole occurs, i.e.~$\mu_L=C\mu_0$. The Landau pole condition in the new theory $\alpha'^{-1}(C\mu_0)=0$ can be related to $\alpha'^{-1}(\mu)$ by using the solution in Eq.~\eqref{eq:RGE-solution} (assuming $\beta_f$ for the beta function above $\mu_0$ and no new thresholds); when combined with Eq.~\eqref{eq:g-shift} this relates $c$, $C$ and the original gauge coupling $\alpha^{-1}$ at the GUT scale:
\begin{align}
2\pi \alpha^{-1}(\mu_0)+\beta_f\log C+\Delta\beta\;\log c&=0.\label{eq:cc-connection}
\end{align}
\item We now turn to our specific case of $\SU(5)$ and assume the full theory has at least the following chiral supermultiplets in the fermionic and Higgs sectors (no messengers, no flavons, we assume GUT breaking only with the $\mathbf{24}$ and $\mathbf{75}$):
\begin{align}
3\times (\mathbf{10}_F\oplus\mathbf{\bar{5}}_F) \oplus (\mathbf{50}_H\oplus\overline{\mathbf{50}}_H\oplus \mathbf{5}_H\oplus\overline{\mathbf{5}}_H)\oplus\mathbf{75}_H\oplus\mathbf{24}_H.
\end{align}
This yields $\beta_f=-57$ (see \cite{Slansky:1981yr,Feger:2012bs,Feger:2019tvk} for tables of Dynkin indices), and assuming $g(\mu_0)\approx 0.7$ and $\Delta\beta=-35$ (we move the $\mathbf{50}_H\oplus\overline{\mathbf{50}}$ a factor $c$ below $\MGUT$), we obtain from Eq.~\eqref{eq:cc-connection} the approximate value
\begin{align}
c\approx 10^2/C^{1.63}.
\end{align}
Even for a nearby Landau pole a factor only $C=10$ above $\MGUT$, we can move the $\mathbf{50}$s only a factor $c\approx 2.35$ below the GUT scale.
\end{enumerate}

\noindent
The above calculation implies that to avoid a Landau pole too close to $\MGUT$, the $\mathbf{50}$ cannot really be much lighter than the GUT scale. A small $m_{50}$ alone thus cannot be used to enhance $\MEFF$ by a large factor in the usual missing partner mechanism.

To avoid problems with perturbativity, we need to modify the missing partner mechanism so that the $2$-$2$ element of $\mathbf{M}_T$ has a richer structure. We can for example introduce representations $\mathbf{24}'_H$ and $\mathbf{75}'_H$ (possibly those that we already have of this dimension, depends on the full model), and two new terms to the superpotential $W_{DT}$ in Eq.~\eqref{eq:mp-W}: 

\begin{align}
W'_{DT}&=W_{DT}+
\lambda_{24}\;\mathbf{50}_H\cdot\langle\mathbf{24}'_H\rangle\cdot\mathbf{\overline{50}}_H+
\lambda_{75}\;\mathbf{50}_H\cdot\langle\mathbf{75}'_H\rangle\cdot\mathbf{\overline{50}}_H.
\end{align}
This generates the triplet mass matrix of the new form
\begin{align}
\mathbf{M}_T&=
\begin{pmatrix}
0&\lambda v\\
\bar{\lambda} v& \tilde{m}_{50}\\
\end{pmatrix},
\end{align}
where 
\begin{align}
\tilde{m}_{50}&=m_{50}+\lambda_{24}\langle\mathbf{24}'_{H}\rangle+\lambda_{75}\langle\mathbf{75}'_{H}\rangle.\label{eq:tilde-m50}
\end{align}
The expression for $\tilde{m}_{50}$ can now have all terms at the GUT scale, but the couplings $\lambda_{24}$ and $\lambda_{75}$ can be tuned such that $\tilde{m}_{50}$ itself for the triplets is small, while keeping all SM representations in the $\mathbf{50}_H\oplus\mathbf{\overline{50}}_H$ other than the triplet heavy. This can be seen from the (relative) CG factors in front of the $3$ operators generating their masses that we computed explicitly in Table~\ref{tab:DT-Clebsch}. We can see from this table that the CG factor pairs for the operators $(\lambda_{24},\lambda_{75})$ in each row are unique; since the mass of any SM irreducible representation in the $\mathbf{50}_H\oplus\mathbf{\overline{50}}_H$ has the form of Eq.~\eqref{eq:tilde-m50}, but with CG factors added to the $\lambda_{24}$ and $\lambda_{75}$ terms, the uniqueness of the CG pairs implies that any one SM irreducible representations can be tuned to be light while keeping the others heavy; we choose  to tune it to a low value for the triplets.

In this modified missing partner mechanism, we achieved what we wanted: it is possible to perform DT-splitting, as well as tune $\MEFF$ to arbitrarily high values. All physical masses of particles in the Higgs sector are around the GUT scale (except the MSSM doublet pair), while precision of the tuning in the parameter $\tilde{m}_{50}$ controls how high the effective triplet mass $\MEFF$ is:
\begin{align}
\MEFF&=-\lambda\bar{\lambda}\,v^2/\tilde{m}_{50}.
\end{align}
For $\lambda\bar{\lambda}v^2\approx (10^{16}\GeV)^2$, and effective triplet mass $|\MEFF|=10^{19}\GeV$ used as a benchmark value in this paper, we require $\tilde{m}_{50}\approx 10^{-3}\MGUT$, i.e.~a relatively small tuning of one part in $10^{3}$, which retains the advantage of the missing partner mechanism over a brute force fine-tuning of a doublet from $\MGUT$ to the EW scale. 

\begin{table}[htb]
\caption{Clebsch factors (up to an overall VEV normalization factor) for all different SM representations in the $\mathbf{50}$ (or $\overline{\mathbf{50}}$) of $\SU(5)$ in the three different operators generating their masses.\label{tab:DT-Clebsch}}
\begin{center}
\vskip -0.5cm
\begin{tabular}{lrrr}
\toprule
$G_{321}$ representation&$\mathbf{50}\cdot \overline{\mathbf{50}}$&$\mathbf{50}\cdot\mathbf{24}\cdot\overline{\mathbf{50}}$&$\mathbf{50}\cdot \mathbf{75}\cdot\overline{\mathbf{50}}$\\
\midrule 
$(\mathbf{1},\mathbf{1},-2)$&$1$&$-12$&$-3$\\[3pt]
$(\mathbf{3},\mathbf{1},-\frac{1}{3})$&$1$&$-2$&$-1$\\[3pt]
$(\mathbf{\bar{3}},\mathbf{2},-\tfrac{7}{6})$&$1$&$-7$&$-1$\\[3pt]
$(\mathbf{\bar{6}},\mathbf{3},-\tfrac{1}{3})$&$1$&$-2$&$1$\\[3pt]
$(\mathbf{6},\mathbf{1},+\tfrac{4}{3})$&$1$&$8$&$-1$\\[3pt]
$(\mathbf{8},\mathbf{2},+\tfrac{1}{2})$&$1$&$3$&$0$\\
\bottomrule
\end{tabular}
\end{center}
\end{table}

\subsection{Double missing partner mechanism \label{sec:double-missing-partner}}

A straightforward alternative to the modified missing partner mechanism discussed in Appendix~\ref{sec:tweak} is the \textit{double missing partner mechanism}, cf.~\cite{Antusch:2014poa}. This also solves the DT splitting problem and provides only one effective triplet mass that can be trans-Planckian.

In this scenario, the representations of the doublets/triplets are doubled: the
DT sector consists of
\begin{align}
(\mathbf{5}_H\oplus\mathbf{\bar{5}}_H\oplus\mathbf{50}_H\oplus\mathbf{\overline{50}}_H)\oplus (\mathbf{5}'_H\oplus\mathbf{\bar{5}}'_H\oplus\mathbf{50}'_H\oplus\mathbf{\overline{50}}'_H),
\end{align}
where the $G_{321}$ doublets $(\mathbf{1},\mathbf{2},\pm 1/2)$ are contained in
\begin{align}
D_1&\subseteq \mathbf{5}_H,& \bar{D}_1&\subseteq \mathbf{\bar{5}}_H,\\
D'_1&\subseteq \mathbf{5}'_H,& \bar{D}'_1&\subseteq \mathbf{\bar{5}}'_H,
\end{align}
and the triplets $T_{I}\sim (\mathbf{3},\mathbf{1},-1/3)$ and anti-triplets $\overline{T}_I\sim (\mathbf{\bar{3}},\mathbf{1},+1/3)$ are in
\begin{align}
T_1&\subseteq \mathbf{5}_H,&\overline{T}_1 &\subseteq \mathbf{\bar{5}}_H,\nonumber\\
T'_1&\subseteq \mathbf{5}'_H,&\overline{T}'_1 &\subseteq \mathbf{\bar{5}}'_H,\nonumber\\
T_2&\subseteq \mathbf{50}_H,&\overline{T}_2 &\subseteq \mathbf{\overline{50}}_H,\nonumber\\
T'_2&\subseteq \mathbf{50}'_H,&\overline{T}'_2 &\subseteq \mathbf{\overline{50}}'_H.
\end{align}
We then arrange the presence of such operators that the mass terms 
\begin{align}
\mathbf{W}\supset 
\begin{pmatrix}
D_{1}\\
D'_{1}\\
\end{pmatrix}^{\!\tr}
\mathbf{M}_{D}
\begin{pmatrix}
\overline{D}_{1}\\
\overline{D}'_{1}\\
\end{pmatrix}
+
\begin{pmatrix}
T_{1}\\
T'_{1}\\
T_{2}\\
T'_{2}\\
\end{pmatrix}^{\!\tr}
\mathbf{M}_{T}
\begin{pmatrix}
\overline{T}_{1}\\ 
\overline{T}'_{1}\\ 
\overline{T}_{2}\\
\overline{T}'_{2}\\ 
\end{pmatrix}+\cdots,
\end{align}
have the following specific form for the mass matrices:
\begin{align}
\mathbf{M}_{D}&=
\begin{pmatrix}
0&0\\
0&\mu'\\
\end{pmatrix}
,&
\mathbf{M}_{T}&=
\begin{pmatrix}
0&0&0&\alpha_1 V\\
0&\mu'&\alpha_2 V&0\\
\alpha_3 V&0&M_{50}&0\\
0&\alpha_4 V&0&M'_{50}\\
\end{pmatrix},\label{eq:double-missing-partner-MT}
\end{align}
where $\alpha_i$ are dimensionless couplings, while $V$ could for example be a VEV $V=\langle \mathbf{75}\rangle$ at the GUT scale (see \cite{Antusch:2014poa} for more details on some realizations). In this set up, we require that only $\mathbf{5}_H\oplus\overline{\mathbf{5}}_H$ (and not those with primes) can couple to the SM fermions in $\mathbf{10}_{Fi}\oplus \overline{\mathbf{5}}_{Fi}$, which yields the effective triplet setup with the effective mass parameter equal to
\begin{align}
\MEFF=((\mathbf{M}_T^{-1})_{11})^{-1}=-\alpha_1 \alpha_2 \alpha_3 \alpha_4 \frac{V^{4}}{\mu'\,M_{50}\,M'_{50}}.
\end{align}
To postpone the appearance of the Landau pole from the large (negative) contributions to the beta function, we can provide the representations $\mathbf{50}\oplus \mathbf{50}'\oplus \overline{\mathbf{50}}\oplus \overline{\mathbf{50}}'$ with large mass terms close to the Planck scale, i.e.~we take the hierarchy 
\begin{align}
\mu' \quad\ll\quad V\approx \MGUT \quad\ll\quad M_{50} \approx M'_{50} \quad\lesssim\quad  M_{Pl}.
\end{align}

We obtain from this setup the physical doublet masses to be $0$ (the MSSM doublet) and $\mu'$, while the physical triplet masses are at scales $\eta^{\pm 1} V$ (two triplets at each scale), where $\eta\equiv V/M_{50}$. This constitutes a successful DT splitting yielding
\begin{align}
\MEFF\approx -\eta^{2}\frac{V^2}{\mu'},
\end{align}
allowing for the effective mass to be large provided that $\mu'$ is small. For \hbox{$V=2\cdot 10^{16}\GeV$,} \hbox{$\eta^{2}\approx 10^{-4}$} (so that $M_{50}$ and $M'_{50}$ are close to the Planck scale) and \hbox{$|\MEFF| \gtrsim 10^{19}\GeV$,} we need \hbox{$|\mu'|\lesssim 4\cdot 10^9\GeV$.}  Note that $\mu'$ is associated to the scale of the second doublet (a rather small representation not irreparably spoiling the unification of gauge couplings), but not to any other particle, since the $\mathbf{5}$s contain beside the doublets only the already considered triplets. For a more precise calculation, such a scenario does require adding an extra doublet to the MSSM at the intermediate scale $\mu'$.

\section{Documentation for the proton decay calculation \label{sec:documentation}}
\subsection{Numerical procedure}
\subsubsection{Definition of the dimension~5 operators\label{appendix:quasiYukawa-definitions}}
The effective superpotential of the dimension~$5$ operators, which are relevant for proton and neutron decay, reads
\begin{align}
\begin{split}
W_5 &= -\tfrac{1}{2}\mathbf{C}^{ijkl}_{5L}\,\epsilon_{\ah\bh\ch}\,(Q_i^\ah\cdot L_j)(Q_k^\bh\cdot Q_l^\ch) \\
&\quad - \mathbf{C}^{ijkl}_{5R}\,\epsilon^{\ah\bh\ch}\,u^c_{i\ah}\,d^c_{j\bh}\,e^c_k\,u^c_{l\ch}\,,
\end{split} \label{eq:documentation_definition_W5}
\end{align}
where $\ah,\bh,\ch$ are $\SU(3)_\text{C}$ indices with $\epsilon_{123}=\epsilon^{123}=1$, and the contraction of $\SU(2)_\text{L}$ indices is given by $\Psi\cdot\Phi=\epsilon_{ab}\,\Psi^a\,\Phi^b$ with $\epsilon_{12}=1$.
\par
The quasi-Yukawa operators and the mass term for the (anti-)triplets are given by
\begin{align}
\begin{split}
W_\text{T} &= \phantom{+} \tfrac{1}{2}(\tilde{\mathbf{Y}}_{qq})_{Iij}\,\epsilon_{\ah\bh\ch}\,Q_i^\ah\cdot Q_j^\bh\, T_I^\ch + (\tilde{\mathbf{Y}}_{eu})_{Iij}\,e^c_i\,u^c_{j\ah}\,T_I^\ah \\
&\quad + (\tilde{\mathbf{Y}}_{ql})_{Jij}\,Q_i^\ah\cdot L_j\,\overline{T}_{J\ah} + (\tilde{\mathbf{Y}}_{ud})_{Jij}\,\epsilon^{\ah\bh\ch}\,u^c_{i\ah}\,d^c_{j\bh}\,\overline{T}_{J\ch} \\
&\quad + (\mathbf{M}_T)_{IJ}\,T_I^\ah\,\overline{T}_{J\ah}\,.
\end{split} \label{eq:documentation_definition_quasiyukawa}
\end{align}
After integrating out the (anti-)triplets, the effective baryon and lepton number violating operators are obtained:
\begin{align}
\begin{split}
W_5 &= -\tfrac{1}{2}(\mathbf{M}_T^{-1})_{IJ}\,(\tilde{\mathbf{Y}}_{ql})_{Iij}\,(\tilde{\mathbf{Y}}_{qq})_{Jkl}\,\epsilon_{\ah\bh\ch}\,(Q_i^\ah\cdot L_j)(Q_k^\bh\cdot Q_l^\ch) \\
&\quad - (\mathbf{M}_T^{-1})_{IJ}\,(\tilde{\mathbf{Y}}_{ud})_{Iij}\,(\tilde{\mathbf{Y}}_{eu})_{Jkl}\,\epsilon^{\ah\bh\ch}\,u^c_{i\ah}\,d^c_{j\bh}\,e^c_k\,u^c_{l\ch} \\
&\quad + ...\,,
\end{split} \label{eq:documentation_effective_superpotential}
\end{align}
where the dots indicate terms which are not relevant for nucleon decay. According to Eq.~\eqref{eq:documentation_definition_W5}, the dimension~$5$ operators at the GUT scale are then calculated as follows:
\begin{align}
\mathbf{C}^{ijkl}_{5L} &=  (\mathbf{M}_T^{-1})_{IJ}\,(\tilde{\mathbf{Y}}_{ql})_{Iij}\,(\tilde{\mathbf{Y}}_{qq})_{Jkl}\,, \label{eq:documentation_dim5ops_def_1}\\
\mathbf{C}^{ijkl}_{5R} &=  (\mathbf{M}_T^{-1})_{IJ}\,(\tilde{\mathbf{Y}}_{ud})_{Iij}\,(\tilde{\mathbf{Y}}_{eu})_{Jkl}\,. \label{eq:documentation_dim5ops_def_2}
\end{align}

\subsubsection{Running of the dimension~5 operators}
The running of the dimension~$5$ operators from the GUT scale to the SUSY scale is calculated in the framework of the MSSM. The $\beta$-functions of the dimension~$5$ operators in the MSSM at $1$-loop in the $\overline{\mathrm{DR}}$ scheme, by using the left-right convention for the Yukawa matrices, are given by
\begingroup
\allowdisplaybreaks
\begin{align}
\begin{split}
\beta(\mathbf{C}_{5L}^{ijkl}) &= \phantom{+}
\mathbf{C}_{5L}^{mjkl} \big(\YD\YD^\ct + \YU\YU^\ct\big)^i{}_m
+ \mathbf{C}_{5L}^{imkl}\big(\YE\YE^\ct + \YNU\YNU^\ct\big)^j{}_m \\
&\quad + \mathbf{C}_{5L}^{ijml} \big(\YD\YD^\ct + \YU\YU^\ct\big)^k{}_m
+ \mathbf{C}_{5L}^{ijkm} \big(\YD\YD^\ct + \YU\YU^\ct\big)^l{}_m \\
&\quad + \mathbf{C}_{5L}^{ijkl} \big(-\tfrac{2}{5}g_1^2 - 6g_2^2 - 8g_3^2\big)\,,
\end{split} \label{eq:documentation_c5Lbeta}\\
\nonumber\\
\begin{split}
\beta(\mathbf{C}_{5R}^{ijkl}) &= \phantom{+}
\mathbf{C}_{5R}^{mjkl} \big(2\YU^\ct\YU\big)_m{}^i
+ \mathbf{C}_{5R}^{imkl}\big(2\YD^\ct\YD\big)_m{}^j \\
&\quad + \mathbf{C}_{5R}^{ijml} \big(2\YE^\ct\YE\big)_m{}^k
+ \mathbf{C}_{5R}^{ijkm} \big(2\YU^\ct\YU\big)_m{}^l \\
&\quad + \mathbf{C}_{5R}^{ijkl} \big(-\tfrac{12}{5}g_1^2 - 8g_3^2\big)\,.
\end{split} \label{eq:documentation_c5Rbeta}
\end{align}
\endgroup
The $\beta$-functions are computed by means of the general formulas in~\cite{Martin:1993zk} for the \hbox{$1$-loop} anomalous dimension matrix ${\gamma_m}^i$, and the non-renormalization theorem of the superpotential~\cite{Wess:1973kz,Iliopoulos:1974zv,Grisaru:1979wc}. We use the GUT normalization for the $\U(1)_\text{Y}$ gauge coupling~$g_1$; the SM normalization can be recovered by $g_1^\text{SM}=\sqrt{3/5}\,g_1$. The RGEs are then given by
\begin{align}
\frac{\mathrm{d}x}{\mathrm{d}\log\mu}=\frac{1}{16\pi^2}\beta(x)\,, \label{eq:documentation_c5running_rge}
\end{align}
where $x\in\{\mathbf{C}_{5L}^{ijkl},\mathbf{C}_{5R}^{ijkl}\}$ and $\mu$ is the renormalization scale.

\subsubsection{Dressing of the dimension~5 operators}
\begin{figure}[htb]
\begin{center}
\caption{An example diagram for the proton decay channel $p\rightarrow K^+\bar{\nu}$. The big black circle represents the effective dimension~$5$ operator, and the dressing of this operator takes place by using a mass insertion of the winos $\wt{W}^\pm$. \label{fig:documentation_dressing_example}}
\includegraphics[scale=0.7]{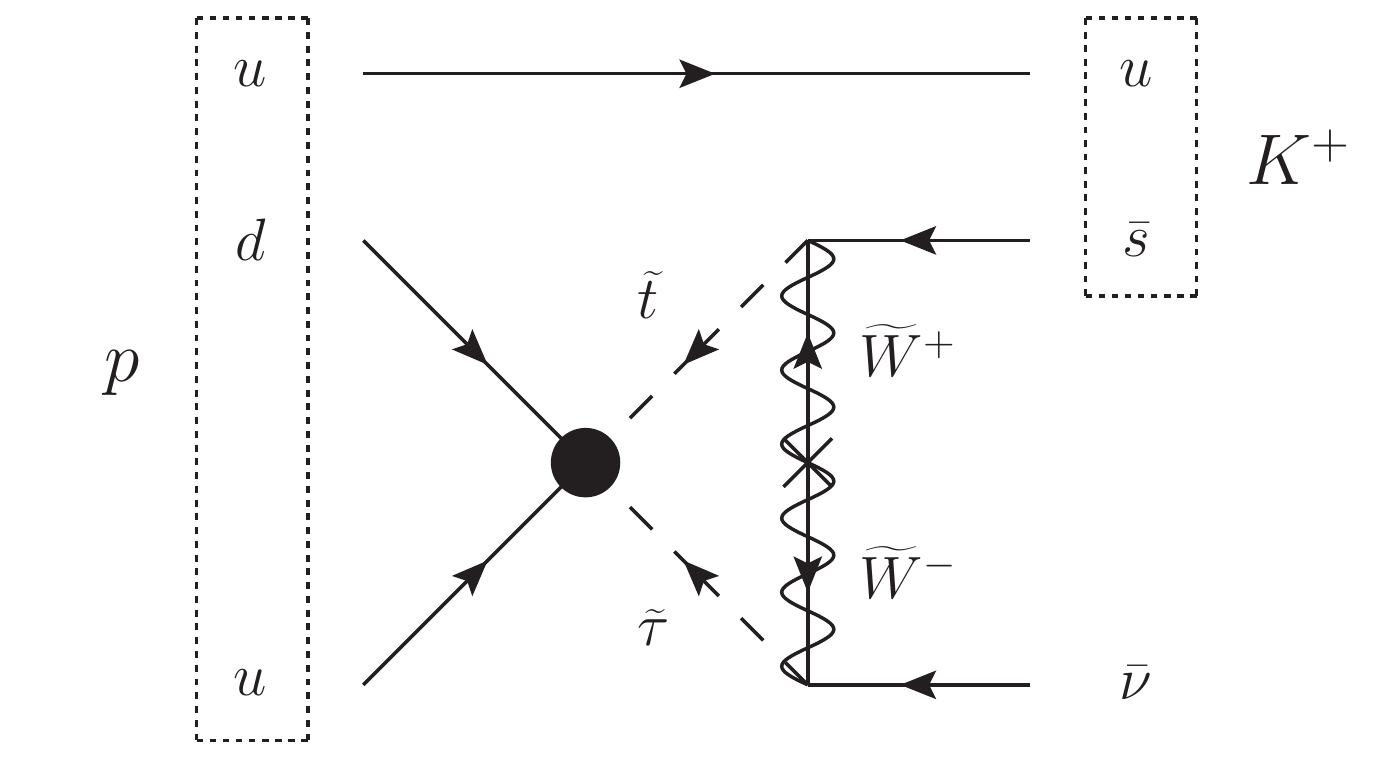}
\end{center}
\end{figure}

The dimension~$6$ four-fermion operators are built at $1$-loop level by dressing the dimension~$5$ operators with gluino, chargino and neutralino exchange diagrams, which takes place at the SUSY scale. An example diagram is shown in Figure~\ref{fig:documentation_dressing_example}. The complete set of formulas to calculate the dressing of the dimension~$5$ operators is stated in~\cite{Goto:1998qg}, in particular Eqs.~(A.1.13)--(A.1.16) and Eqs.~(A.2.8)--(A.2.20). Furthermore, in~\cite{Bajc:2002bv} all relevant diagrams for nucleon decay from dimension~$5$ operators can be found.
\par
Beside the dimension~$5$ operators, the calculation of the resulting dimension~$6$ operators involves the rotation matrices of the sfermions, charginos and neutralinos, as well as the mass eigenvalues of the sparticles. It is convenient to use the standardized SLHA~\cite{Skands:2003cj} and SLHA2~\cite{Allanach:2008qq} conventions to specify these quantities. In Section~\ref{sec:documentation_slha_conventions} the translation of the conventions used in~\cite{Goto:1998qg} into the SLHA conventions is stated.
\par
Apart from integrating out the sparticles at the SUSY scale, we also switch to the EW symmetry broken phase of the SM. In the following $u_i$, $d_i$ and $e_i$ $(i\in\{1,2,3\})$ represent the mass eigenstates of the up-type quarks, the down-type quarks and the charged leptons. In contrast, the $\nu_i$ represent the neutrinos in the interaction (flavor) basis. The additional label $L$ or $R$ indicates whether a state is a left- or a right-handed Weyl spinor. The dimension~$6$ operators $\wt{\mathbf{C}}$ (at the Lagrangian level), which are relevant for the calculation of proton decay, include only mass eigenstates of the fermions which are lighter than the nucleons and are given by
\begin{align}
\begin{split}
\mathcal{L}_{\slashed{B}} &= \frac{1}{16\pi^2}\epsilon_{\ah\bh\ch}\Big( \wt{\mathbf{C}}_{LL}(udue)^{ik}(u_{L}^\ah d_{Li}^\bh)(u_{L}^\ch e_{Lk}^\phant) + \wt{\mathbf{C}}_{RL}(udue)^{ik}(u_{R}^\ah d_{Ri}^\bh)(u_{L}^\ch e_{Lk}^\phant) \\
&\qquad\qquad\;\; + \wt{\mathbf{C}}_{LR}(udue)^{ik}(u_{L}^\ah d_{Li}^\bh)(u_{R}^\ch e_{Rk}^\phant) + \wt{\mathbf{C}}_{RR}(udue)^{ik}(u_{R}^\ah d_{Ri}^\bh)(u_{R}^\ch e_{Rk}^\phant) \\
&\qquad\qquad\;\; + \wt{\mathbf{C}}_{LL}(udd\nu)^{ijk}(u_{L}^\ah d_{Li}^\bh)(d_{Lj}^\ch \nu_{Lk}^\phant) + \wt{\mathbf{C}}_{RL}(udd\nu)^{ijk}(u_{R}^\ah d_{Ri}^\bh)(d_{Lj}^\ch \nu_{Lk}^\phant) \\
&\qquad\qquad\;\; + \tfrac{1}{2}\wt{\mathbf{C}}_{RL}(ddu\nu)^{ijk}(d_{Ri}^\ah d_{Rj}^\bh)(u_{L}^\ch \nu_{Lk}^\phant) \Big) + h.c.\,.
\end{split} \label{eq:documentation_c6definition}
\end{align}
Since only the lightest up-type quark $u_1$ is lighter than the proton and the neutron, only this state is present in Eq.~\eqref{eq:documentation_c6definition} and the label $i$ is neglected.\footnote{Note that in the calculation of the decay widths, the index $i$ in $d_i$ and $e_i$ is restricted to the values $1,2$, because the heaviest down-type quark and charged lepton $(i=3)$ are heavier than the two nucleons. In contrast, all three states $\nu_i$ are present.}

\subsubsection{Running of the dimension~6 operators}
The running of the dimension~$6$ operators from the SUSY scale to the mass scale of the nucleons is calculated in the framework of the SM. Following~\cite{Nihei:1994tx}, the main contribution to the running of the dimension~$6$ operators comes from the strong gauge coupling $g_3$. At $1$-loop the $\beta$-functions in the $\overline{\mathrm{MS}}$ scheme of the dimension~$6$ operators $\wt{\mathbf{C}}$ from Eq.~\eqref{eq:documentation_c6definition}, written in the general form $\wt{\mathbf{C}}^{ijkl}$, and of the gauge coupling $g_3$ are given by
\begin{align}
\beta(\wt{\mathbf{C}}^{ijkl}) &= -4\,g_3^2\,\wt{\mathbf{C}}^{ijkl}\,, \\
\beta(g_3) &= \Big(-11+\frac{2}{3}N_F\Big)g_3^3\,,
\end{align}
where $N_F$ is the number of quarks whose masses are below the renormalization scale $\mu$. The RGEs have the form as specified in Eq.~\eqref{eq:documentation_c5running_rge}. The running of the dimension~$6$ operators just corresponds to an overall scaling factor, which is also referred to as the long range effect on the effective operators $\wt{\mathbf{C}}$.

\subsubsection{Partial decay widths}
\label{sec:protondecay_numeric_decaywidths}
The formula for the calculation of the partial decay widths of a nucleon $B_i$ with a meson $M_j$ and a lepton $l_k$ in the final state is taken from~\cite{Goto:1998qg} and is given by
\begin{align}
\Gamma(B_i \rightarrow M_j\,l_k) &= \frac{m_i}{32\pi}\, \Big(1-\frac{m^2_j}{m^2_i}\Big)^2\, \frac{1}{f_\pi^2}\, \Big(\big|A_L^{ijk}\big|^2 + \big|A_R^{ijk}\big|^2\Big)\,,
\end{align}
where $m_i$ and $m_j$ are the masses of the nucleon and meson, respectively. Moreover, $f_\pi=0.130\GeV$ (cf.\ PDG~\cite{Tanabashi:2018oca}) is the pion decay constant.
\par
The amplitudes $A_L^{ijk}$ and $A_R^{ijk}$ for different decay channels are calculated by using the dimension~$6$ operators at the nucleon mass scale, and are specified in Table~1 in~\cite{Goto:1998qg}. In that table, the nucleon mass $m_N$ is taken as the average of the proton mass $m_p$ and the neutron mass $m_n$, and the baryon mass $m_{B'}$ is given by the average of $m_\Lambda$ and $m_\Sigma$. The additional constants with the corresponding (approximate) values are the constants from hyperon decay $F\approx0.463$, $D\approx0.804$ (cf.~\cite{Cabibbo:2003cu,Nath:2006ut}), and the parameters for the proton decay matrix element $\alpha\approx0.0090\GeV^3$, $\beta\approx0.0096\GeV^3$ (cf.~\cite{Tsutsui:2004qc,Nath:2006ut}).
\par
The following decay channels for the proton $p$ and the neutron $n$ are considered in the calculation of the partial decay widths:
\begin{equation}
\begin{aligned}
&\bullet \text{ proton: } &&p\rightarrow \pi^0\,e_k^+,\quad p\rightarrow \eta^0\,e_k^+,\quad p\rightarrow K^0\,e_k^+,\quad p\rightarrow \pi^+\,\bar{\nu}_k,\quad p\rightarrow K^+\,\bar{\nu}_k, \\\addlinespace[0.3cm]
&\bullet \text{ neutron: } &&n\rightarrow \pi^-\,e_k^+,\quad n\rightarrow \pi^0\,\bar{\nu}_k,\quad n\rightarrow \eta^0\,\bar{\nu}_k,\quad n\rightarrow K^0\,\bar{\nu}_k.
\end{aligned} \label{eq:documentation_decaywidths_channels}
\end{equation}
As discussed above, $k\in\{1,2\}$ for $e_k^+$, since the heaviest charged lepton $\tau$ corresponding to $k=3$ is heavier than the nucleons, and $k\in\{1,2,3\}$ for $\bar{\nu}_k$.

\subsection{Matching with SLHA conventions}
\label{sec:documentation_slha_conventions}
In the following the relation between the conventions in~\cite{Goto:1998qg} (later referred to as GN), which are used in the dressing of the dimension~$5$ operators, and the standardized set of conventions in SLHA~\cite{Skands:2003cj} and SLHA2~\cite{Allanach:2008qq} are specified.
\begin{itemize}
\item The Yukawa matrices are related as follows:
\begin{center}
\myArrayStretch
\begin{tabular}{llcl}
\toprule
GN & SLHA & \multicolumn{2}{l}{our notation}\\
\midrule
$f_U$ & $Y_U$ & = & $f_U$ \\
$f_D$ & $Y_D$ & = & $-f_D$ \\
$f_L$ & $Y_E$ & = & $f_L^\tr$ \\
\bottomrule
\end{tabular}
\myArrayStretchReset
\end{center}
where the Yukawa matrices are given in GN: Eq.~(A.1.1) and SLHA: Eq.~(3). In particular, GN uses the LR convention in the up- and down-quark sector ($f_U$ and $f_D$), and the right-left (RL) convention in the charged lepton sector $(f_L)$. On the other hand, SLHA uses the LR convention in all sectors.

\item Both GN and SLHA use the SM normalization for the gauge coupling $g_1$ of $\U(1)_\text{Y}$, and they also use the same convention for the $\mu$-term.

\item In the EW symmetry broken phase the mass matrices of the sfermions are written in a basis where the sfermions are aligned with their SM superpartners.
\begin{itemize}
\item GN uses the basis where the mass matrices $M_U=M_U^\text{diag}$ (up-quarks) and $M_L=M_L^\text{diag}$ (charged leptons) are diagonal, and the mass matrices $M_D = \Vckm^*\,M_D^\text{diag}$ (down-quarks) and $M_\nu=\Vpmns^*\,M_\nu^\text{diag}\,\Vpmns^\tr$ (neutrinos) are rotated by the CKM and PMNS matrix, respectively ($u_L$ and $d_L$ as well as $e_L$ and $\nu_L$ are assumed to form doublets under rotations).
\item SLHA uses the basis where all fermion mass matrices are diagonal, which corresponds to the super-CKM/PMNS basis.
\end{itemize}
The relation between the sfermion, the chargino and the neutralino mass matrices in GN and SLHA are then the following:
\begin{center}
\myArrayStretch
\begin{tabular}{llcll}
\toprule
GN & SLHA \\
\midrule
$\cM^2_{\wt{u}}$ & $(\cM^2_{\wt{u}})_\text{sCKM}$ & $=$ & $(\cM^2_{\wt{u}})^\textsf{T}$ \\
$\cM^2_{\wt{d}}$ & $(\cM^2_{\wt{d}})_\text{sCKM}$ & $=$ & $U_d (\cM^2_{\wt{d}})^\textsf{T} U_d^\dagger$ & with $U_d = \left(\begin{smallmatrix} V^\dagger_\text{CKM} & 0\\ 0 & \mathbb{1} \end{smallmatrix}\right)$ \\
$\cM^2_{\wt{e}}$ & $(\cM^2_{\wt{e}})_\text{sPMNS}$ & $=$ & $\cM^2_{\wt{e}}$ \\
$\cM^2_{\wt{\nu}}$ & $(\cM^2_{\wt{\nu}})_\text{sPMNS}$ & $=$ & $U_\nu (\cM^2_{\wt{\nu}})^\textsf{T} U_\nu^\dagger$ & with $U_\nu = \Vpmns^\ct$ \\
$\cM_C$ & $\cM_{\wt{\psi}^+}$ & $=$ & $-\cM_C$ \\
$\cM_N$ & $\cM_{\wt{\psi}^0}$ & $=$ & $\cM_N$ \\
\bottomrule
\end{tabular}
\myArrayStretchReset
\end{center}
The mass matrices of the sfermions are given in GN: Eqs.~(A.1.6)--(A.1.9) and SLHA2: Eq.~(11), (12), (24) and (25). In addition, the mass matrices of the charginos and neutralinos are stated in GN: Eq.~(A.1.11) and SLHA: Eq.~(21) and (22).
\par
The mass matrices of the squarks and charged sleptons $\wt{f}$ $(f\in\{u,d,e\})$ are $6\times6$-dimensional, written in the basis $(\wt{f}_{L1},\wt{f}_{L2},\wt{f}_{L3},\wt{f}_{R1},\wt{f}_{R2},\wt{f}_{R3})$, where $\wt{f}_{L,Ri}$ is the superpartner of $f_{L,Ri}$, i.e.\
\begin{align}
\cM^2_{\wt{f}} &= \begin{pmatrix}
\cM^2_{LL} & \cM^2_{LR} \\
\cM^2_{RL} & \cM^2_{RR}
\end{pmatrix},
\end{align}
where each entry represents a $3\times3$-block. In contrast, the mass matrix $\cM^2_{\wt{\nu}}\equiv\cM^2_{LL}$ of the sneutrinos is only $3\times3$-dimensional, since only left-handed neutrinos are present.
\par
Moreover, the $2\times2$-dimensional mass matrix $\cM_C$ of the charginos is written in the basis $(\wt{W}^\pm,\wt{H}^\pm_u)$, and for the $4\times4$-dimensional mass matrix $\cM_N$ of the neutralinos the basis $(\wt{B},\wt{W}^0,\wt{H}^0_d,\wt{H}^0_u)$ is used.

\item The definitions and relations of the sparticle rotation matrices in GN and SLHA are given by:
\begin{center}
\myArrayStretch
\begin{tabular}{llcl}
\toprule
GN & SLHA & & Relation \\
\midrule
$\wt{U}_U (\cM^2_{\wt{u}})^\textsf{T} \wt{U}_U^\dagger = \text{diag}$ & $R_u (\cM^2_{\wt{u}})_\text{sCKM} R_u^\dagger = \text{diag}$ & $\Rightarrow$ & $\wt{U}_U = R_u$\\
$\wt{U}_D (\cM^2_{\wt{d}})^\textsf{T} \wt{U}_D^\dagger = \text{diag}$ & $R_d (\cM^2_{\wt{d}})_\text{sCKM} R_d^\dagger = \text{diag}$ & $\Rightarrow$ & $\wt{U}_D = R_d U_d$\\
$\wt{U}_L^\dagger \cM^2_{\wt{e}} \wt{U}_L = \text{diag}$ & $R_e (\cM^2_{\wt{e}})_\text{sPMNS} R_e^\dagger = \text{diag}$ & $\Rightarrow$ & $\wt{U}_L = R_e^\dagger$\\
$\wt{U}_N^\dagger \cM^2_{\wt{\nu}} \wt{U}_N = \text{diag}$ & $R_\nu (\cM^2_{\wt{\nu}})_\text{sPMNS} R_\nu^\dagger = \text{diag}$ & $\Rightarrow$ & $\wt{U}_N = (R_\nu U_\nu)^\dagger$\\
$-U_{-}^\dagger \cM_C U_{+} = \text{diag}$ & $U (\cM_C)_\text{SLHA} V^\textsf{T} = \text{diag}$ & $\Rightarrow$ & $U_{-} = U^\dagger$, $U_{+} = V^\textsf{T}$\\
$U_N^\textsf{T} \cM_N U_N = \text{diag}$ & $N^* (\cM_N)_\text{SLHA} N^\dagger = \text{diag}$ & $\Rightarrow$ & $U_N = N^\dagger$\\
\bottomrule
\end{tabular}
\myArrayStretchReset
\end{center}
where ``$\diag$'' indicates a diagonal matrix with real, positive entries. The definitions for the sfermion rotation matrices are taken from GN: Eq.~(A.1.10) and SLHA2: Eqs.~(28)--(31), whereas the definitions for the chargino and neutralino rotation matrices are stated in GN: Eq.~(A.1.12) and SLHA: Eq.~(12) and (15).
\end{itemize}

\section{User's guide to Mathematica packages \texttt{SusyTCProton} and \texttt{ProtonDecay} \label{sec:user_guide}}
This section provides a guide for the usage of the Mathematica packages \texttt{SusyTCProton.m} and \texttt{ProtonDecay.m}. Visit the web page \url{http://particlesandcosmology.unibas.ch/downloads/protondecay.html} to obtain the packages. The zip file \texttt{ProtonDecay.zip} there contains the \texttt{README.md} file with installation instructions, the \texttt{.m} package file and a directory containing some example notebooks for evaluation. For \texttt{SusyTCProton}, follow the link on the web page specified above. Mathematica package requirements: \texttt{REAP} for \texttt{SusyTCProton}, none for \texttt{ProtonDecay}.

The documentation below uses the typewriter font to specify functions and variables. In particular, the dimension~$5$ operators $\mathbf{C}^{ijkl}_{5L}$ and $\mathbf{C}^{ijkl}_{5R}$ from Eq.~\eqref{eq:documentation_definition_W5} are labeled as \texttt{CL5qlqq} and \texttt{CR5udeu}, respectively.

\subsection{\texttt{SusyTCProton.m}}
\label{sec:user_guide_SusyTCProton}
The package \texttt{SusyTCProton.m} has the same functionality as the package \texttt{SusyTC.m} from~\cite{Antusch:2015nwi}, with the following extensions/modifications: 
\begin{itemize}
\item \texttt{RGEGetSolution}:\\
This function has the same functionality as described in~\cite{Antusch:2015nwi}, where in addition the values of the dimension~$5$ operators \texttt{RGECL5qlqq} and \texttt{RGECR5udeu} can be output at a given energy scale.

\item \texttt{STCGetProtonDecayOperators[]}:\\
Returns a list of replacement rules for the dimension~$5$ operators in the flavour basis and in the SCKM basis (flavor indices in all sectors are rotated so that the Yukawa matrices are diagonal, see e.g.~\cite{Antusch:2015nwi} for SCKM definition):
	\begin{itemize}
	\item \texttt{"FlavourBasis[CL5qlqq]"}
	\item \texttt{"FlavourBasis[CR5udeu]"}
	\item \texttt{"SCKMBasis[CL5qlqq]"}
	\item \texttt{"SCKMBasis[CR5udeu]"}
	\end{itemize}

\item \texttt{STCGetSUSYSpectrum[]}, \texttt{STCGetSCKMValues[]}, \texttt{STCGetOneLoopValues[]}:\\
These functions are the same as described in~\cite{Antusch:2015nwi}.

\item \texttt{STCGetInternalValues[]}:\\
Returns a list of replacement rules of the quantities which are internally used for the calculation of the threshold corrections and the sparticle spectrum, i.e.~the combined results from \texttt{STCGetSCKMValues[]}, \texttt{STCGetSUSYSpectrum[]} and \texttt{STCGetProtonDecayOperators[]}, with $1$-loop corrected parameters from \texttt{STCGetOneLoopValues[]} replacing tree-level ones if available.
\end{itemize}

\subsection{\texttt{ProtonDecay.m}}
\label{sec:user_guide_ProtonDecay}
The package \texttt{ProtonDecay.m} contains functions to calculate \hbox{$D=5$} and \hbox{$D=6$} nucleon decay according to the considerations in Section~\ref{sec:documentation}.
\begin{itemize}
\item \texttt{RunDim5OperatorsToMSUSY[\{Y,MR,g,C5\},scales]}:\\
Calculates the running of the dimension~$5$ operators from the GUT scale to the SUSY scale at $1$-loop, according to Eq.~\eqref{eq:documentation_c5Lbeta} and \eqref{eq:documentation_c5Rbeta}, and the use of beta functions of the MSSM superpotential parameters therein (see e.g.~\cite{Antusch:2015nwi} for their RGE). The input is specified at the GUT scale and has the following form:
	\begin{itemize}
	\item \texttt{Y = \{Yu,Yd,Ye,Ynu\}}: Yukawa matrices using left-right (SLHA) convention,
	\item \texttt{MR}: right-handed neutrino mass matrix,
	\item \texttt{g = \{g1,g2,g3\}}: gauge couplings (GUT normalization for \texttt{g1} is used),
	\item \texttt{C5 = \{C5Lqlqq,C5Rudeu\}}: dimension~$5$ operators in the SCKM basis,
	\item \texttt{scales = \{MGUT,MSUSY\}}: GUT scale and SUSY scale.
	\end{itemize}
The output is given at the SUSY scale and has the form \texttt{\{Y,MR,g,C5\}}, where the same definitions as for the input are used.

\item \texttt{ProtonDecayWidths}:\\
Calculates the partial decay widths of the proton and neutron for the decay channels listed in Eq.~\eqref{eq:documentation_decaywidths_channels}. To calculate \hbox{$D=5$} nucleon decay (optionally combined with \hbox{$D=6$} nucleon decay), the function accepts two different inputs at the SUSY scale, where the input of the dimension~$6$ operators
\begin{align}
\texttt{C6 = \{CLLudue,CRLudue,CLRudue,CRRudue,CLLuddn,CRLuddn,CRLddun\}}
\end{align}
is optional (cf.\ Eq.~\eqref{eq:documentation_c6definition} for the definitions, where $\wt{\mathbf{C}}$ are tensors of dimension $3\times3$ or $3\times3\times3$, i.e. $i,(j),k \in \{1,2,3\}$):
	\begin{itemize}
	\item \texttt{ProtonDecayWidths[STCInternalValues,C6\textit{(optional)}]}, where
		\begin{itemize}
		\item[$\bullet$] \texttt{STCInternalValues}: output of the function \texttt{STCGetInternalValues[]} from the package \texttt{SusyTCProton},
		\item[$\bullet$] \texttt{C6}: dimension~$6$ operators at $M_\text{SUSY}$.
		\end{itemize}
	\item \texttt{ProtonDecayWidths[pathSLHA2File,C5,C6\textit{(optional)}]}, where 
		\begin{itemize}
		\item[$\bullet$] \texttt{pathSLHA2File}: path to a SLHA2 output file,
		\item[$\bullet$] \texttt{C5 = \{C5Lqlqq,C5Rudeu\}}: dimension~$5$ operators,
		\item[$\bullet$] \texttt{C6}: dimension~$6$ operators at $M_\text{SUSY}$.
		\end{itemize}
	\end{itemize}
If only \hbox{$D=6$} nucleon decay is calculated, the function accepts the following input at an input mass scale $M_\text{In}$ (typically the EW scale):
	\begin{itemize}
	\item \texttt{ProtonDecayWidths[MIn,g3,C6]}, where 
		\begin{itemize}	
		\item[$\bullet$] \texttt{MIn}: input mass scale $M_\text{In}$,
		\item[$\bullet$] \texttt{g3}: gauge coupling of $\SU(3)_\text{C}$ at $M_\text{In}$,
		\item[$\bullet$] \texttt{C6}: dimension~$6$ operators at $M_\text{In}$.
		\end{itemize}
	\end{itemize}
The function has several option values which appear in the calculation of the partial decay widths (see Sec.~\ref{sec:protondecay_numeric_decaywidths}):
	\begin{itemize}
	\item \texttt{Options[ProtonDecayWidths]}, where
		\begin{itemize}
		\item[$\bullet$] \texttt{"f$\pi$" $\rightarrow$ 0.13}: pion decay constant (in GeV),
		\item[$\bullet$] \texttt{"F" $\rightarrow$ 0.463}: constant from hyperon decay,
		\item[$\bullet$] \texttt{"D" $\rightarrow$ 0.804}: ---{\tt "}---,
		\item[$\bullet$] \texttt{"$\alpha$p" $\rightarrow$ 0.0090}: parameter for the proton decay matrix element,
		\item[$\bullet$] \texttt{"$\beta$p" $\rightarrow$ 0.0096}: ---{\tt "}---.
		\end{itemize}
	\end{itemize}
The output of the function consists of a list with elements of the form \texttt{\{label,decayWidths\}}, where \texttt{label} specifies the decay channel (e.g.\ $p \rightarrow \pi^0\,e^+_k$). Furthermore, \texttt{decayWidths} is the list of the corresponding decay widths in GeV at the nucleon mass scale, where $k\in\{1,2\}$ if the lepton in the final state is a charged lepton, and $k\in\{1,2,3\}$ if the lepton is a neutrino. Note that the charged leptons are considered in the mass eigenbasis, whereas the neutrinos are considered in the interaction (flavor) basis.
\end{itemize}


\end{document}